\documentclass[a4paper,11pt]{article}
\usepackage{jheppub}
\usepackage[T1]{fontenc}
\usepackage[utf8]{inputenc}
\usepackage{amsmath,amssymb}
\usepackage{braket}
\usepackage{graphicx}
\usepackage[dvipsnames,svgnames,x11names]{xcolor}
\usepackage{hyperref}
\usepackage{subfigure}
\usepackage{mathtools}
\usepackage{ulem}

\DeclareMathOperator{\Sch}{Sch}
\DeclareMathOperator{\sech}{sech}
\DeclareMathOperator{\csch}{csch}

\newcommand{\Rmnum}[1]{\expandafter\@slowromancap\romannumeral #1@}

\newcommand{\nn}{\nonumber}

\newcommand{\be}{\begin{equation}}
\newcommand{\ee}{\end{equation}}

\newcommand{\lb}{\left (}
\newcommand{\rb}{\right )}

\renewcommand{\i}{\iota}		

\newcommand{\s}{\sigma}

\newcommand{\R}{\mathbb{R}}

\newcommand{\cQ}{\mathcal{Q}}

\begin{document}


\title{Generalized Clausius inequalities and entanglement production in holographic two-dimensional CFTs}
\author[1]{Tanay Kibe,}
\emailAdd{tanay.kibe@wits.ac.za}
\author[2]{Ayan Mukhopadhyay,}
\emailAdd{ayan.mukhopadhyay@pucv.cl}
\author[1]{and Pratik Roy}
\emailAdd{roy.pratik92@gmail.com }
\affiliation[1]{National Institute for Theoretical and Computational Sciences,
School of Physics\\ and Mandelstam Institute for Theoretical Physics,
University of the Witwatersrand,\\
Wits, 2050, South Africa}
\affiliation[2]{Instituto de F\'{\i}sica, Pontificia Universidad Cat\'{o}lica de Valpara\'{\i}so,
Avenida Universidad 330, Valpara\'{\i}so, Chile.}

\date{\today}

\abstract{Utilizing quantum information theory, it has been shown that irreversible entropy production is bounded from both below and above in physical processes. Both these bounds are positive and generalize the Clausius inequality. Such bounds are, however, obtained from distance measures in the space of states, which are hard to define and compute in quantum field theories. We show that the quantum null energy condition (QNEC) can be utilized to obtain both lower and upper bounds on irreversible entropy production for quenches leading to transitions between thermal states carrying uniform momentum density in two dimensional holographic conformal field theories. We achieve this by refining earlier methods and developing an algebraic procedure for determining HRT surfaces in arbitrary Ba\~nados-Vaidya geometries which are dual to quenches involving transitions between general quantum equilibrium states (e.g. thermal states) where the QNEC is saturated. We also discuss results for the growth and thermalization of entanglement entropy for arbitrary initial and final temperatures and momentum densities. The rate of quadratic growth of entanglement just after the quench depends only on the change in the energy density and is independent of the entangling length. For sufficiently large entangling lengths, the entanglement tsunami phenomenon can be established. Finally, we study recovery of the initial state from the evolving entanglement entropy and argue that the Renyi entropies should give us a refined understanding of scrambling of quantum information.}

\maketitle

\section{Introduction}\label{sec:Introduction}

Irreversible entropy production is associated with any physical process. Recently, a diverse array of methods from non-equilibrium statistical mechanics and quantum information theory \cite{PhysRevLett.105.170402,PhysRevLett.107.140404,Van_2021,RevModPhys.93.035008} have been used for the purpose of understanding irreversible entropy production in processes which take place over a finite time. The methods of quantum thermodynamics particularly have shown that irreversible entropy production is bounded both from above and below in any such process. Both the lower and upper bounds on irreversible entropy production are positive, so that the classical Clausius inequality is only refined in a quantum evolution. However, these bounds have been best understood so far in quantum systems with a finite dimensional Hilbert space. 

In quantum thermodynamics, irreversible entropy production is usually defined in terms of a relative entropy. As for instance, in a unitary evolution with a time-dependent Hamiltonian, the irreversible entropy produced over a given time interval $0\leq t\leq \tau$ is the relative entropy between the final state obtained by this finite time evolution and the thermal state defined via the instantaneous Hamiltonian at the time $\tau$ (see \cite{PhysRevLett.105.170402,RevModPhys.93.035008} for a derivation).\footnote{The irreversible entropy production is close to zero only if the process is almost reversible, i.e. the states at all times are always close to the thermal state defined via the instantaneous Hamiltonians.} A lower bound on the irreversible entropy production can be obtained from the lower bound on relative entropy in terms of distance measures (metrics) in the space of states \cite{PhysRevLett.105.170402}. The upper bound on the irreversible entropy production can be similarly derived \cite{PhysRevLett.105.170402} from the quantum speed limit \cite{PhysRevLett.46.623} stated in terms of a distance measure between the state at a given time and the thermal state defined via the instantaneous Hamiltonian. However, such distance measures are typically hard to both define and compute in quantum field theories. In \cite{Kibe:2021qjy}, it was first shown that the quantum null energy condition (QNEC) can be used to obtain lower and upper bounds on the irreversible entropy production in a relativistic quantum field theory, particularly in quenches in holographic two-dimensional conformal field theories (CFTs) \footnote{Possible bounds on the rate of growth of entanglement entropy in 1+1-D CFTs following quantum quenches, as implications of the QNEC, were discussed in \cite{Mezei:2019sla}. Our results on lower and upper bounds for irreversible entropy production also produce \textit{both} lower and upper bounds on the rate of growth of entanglement entropy following quantum quenches. }. 

The quantum null energy condition (QNEC) in $\mathbb{R}^{1,1}$ states that the inequalities
\begin{equation}\label{Eq:Qpm-def}
    \mathcal{Q}_\pm \coloneq 2 \pi \langle t_{\pm\pm}\rangle - \left(\partial_\pm^2 S_{\rm EE} - \frac{6}{c}\left(\partial_\pm S_{\rm EE}\right)^2\right) \geq 0
\end{equation}
should be satisfied in any physical state in a two-dimensional (1+1-D) CFT with central charge $c$ \cite{Wall:2011kb}. Above, $\langle t_{\pm\pm}\rangle$ is the expectation value of the two non-vanishing null components of the energy momentum tensor at an arbitrary spacetime point $p$ with $+$ ($-$) denoting the right (left) moving future directed null directions, respectively; $S_{\rm EE}$ is the entanglement entropy of any interval of arbitrary length ending at $p$, and $\partial_\pm$ is the rate of change of the entanglement entropy of the interval when the point $p$ is displaced along the corresponding null directions with the other endpoint held fixed. QNEC has been proven for any 1+1-D CFT \cite{Balakrishnan:2017bjg}.\footnote{Originally, it was proposed that a general version of QNEC should hold in relativistic quantum field theories in arbitrary dimensions as it is a necessary condition for the generalized second law \cite{Bekenstein:1974ax} to be satisfied in semi-classical black hole spacetimes harboring quantum fields (just like the classical null energy condition is necessary for the second law of black hole mechanics to hold in spacetimes with classical matter and radiation) \cite{Bousso_2016}. QNEC has now been proven to hold in any relativistic local quantum field theory \cite{Ceyhan:2018zfg}. } 

The quantum null energy condition is saturated (i.e. $\mathcal{Q}_\pm =0$) at all points of spacetime in \textit{quantum equilibrium} states \cite{Ecker:2019ocp}, which are dual to three dimensional (locally AdS$_3$) Ba\~nados geometries \cite{Banados:1998gg} in holographic 1+1-D CFTs at large $c$. A special class of these states are the stationary Ba\~nados-Teitelboim-Zanelli (BTZ) black holes \cite{Banados:1992wn,Banados:1992gq} that are dual to thermal states carrying momentum (with uniform momentum density). Ba\~nados geometries can represent states which are unitary transformations of the thermal rotating states or the vacuum (and are also within the identity block).\footnote{Ba\~nados geometries can represent more exotic possibilities such as disjoint intervals at the boundary which are in different asymptotic regions of the bulk geometry \cite{Sheikh-Jabbari:2016unm}.}

We can thus readily motivate a systematic study of which transitions between quantum equilibrium states are physically realizable especially via a quench involving instantaneous and (in)homogeneous transfer of energy and momentum from an infinite memoryless bath.\footnote{{When a system is in contact with an infinite bath, it is expected to attain thermal equilibrium without changing the bath. When the bath is also memomoryless, the auto-correlation functions of the bath vanish, and the system is expected to undergo  unitary evolution with a time-dependent Hamiltonian. Note that for an isolated system, in contrast, the Hamiltonian is independent of time.}} Note that such a quench is a unitary process in which the Hamiltonian changes rapidly in time. Given that quantum equilibrium states are realizable via unitary transformations of thermal states carrying momenta, they can be used to encode quantum information at finite temperature.  Therefore, a study of the feasibility of such transitions between quantum equilibrium states can lead to the understanding of new principles for the construction of quantum memory and gates operating at finite temperature and in which the \textit{errors} occurring over microscopic time scales can be bounded in a controlled way. As for instance, in \cite{Banerjee:2022dgv}, it was argued that certain dense non-isometric encodings can be fully tolerant to fast deletion at ambient room temperature due to the quantum correction to the Landauer bound.

The Clausius inequality implies that a transition between two thermal states carrying momentum is possible only if the final temperature is larger than the initial temperature, i.e. if irreversible entropy production is positive. In holography, such a transition via a quench can be described via a BTZ-Vaidya geometry in which the input of energy and momentum from an infinite memoryless bath leads to the formation of a null shock. We can expect that quantum thermodynamics should imply finite and positive lower and upper bounds on the production of irreversible entropy for a given amount of energy injection in a quench. Using thermodynamic identities (see Sec.~\ref{Sec:qnechom-quenches}), one can then state that for a fixed change of temperature in a such a quench, there should be positive lower and upper bounds on the production of entropy generalizing the Clausius inequality. For more general transitions between quantum equilibrium states, one can similarly expect that for a fixed amount of total energy injection, there should be lower and upper bounds on how much the final state can differ from the initial one. 

\subsection{Summary of results}

In this work, we first develop an algebraic method of computing entanglement production and studying the validity of the quantum null energy condition in quenches between arbitrary quantum equilibrium states in holographic CFTs. We have refined the previous methods in \cite{Kibe:2021qjy} so that the final state can be largely different from the initial state. Furthermore, we produce exact analytic results pertaining to the growth and thermalization of the entanglement entropy in quenches leading to a transition between \textit{arbitrary} thermal states carrying uniform momentum density (considering that both the temperature and the momentum density can change due to the quench). For an interval of length $l$, thermalization occurs exactly after a time interval $l/2$. We also reproduce many results in the literature related to transitions from the vacuum to the thermal state. Interestingly, we find that some of the results in \cite{Kibe:2021qjy} are also valid beyond the assumed approximations.

We find that just after the quench, the entanglement entropy grows quadratically at a rate which is determined only by the change in the (uniform) energy density and which is independent of the entangling length. We also identify a universal regime for a large interval in which thermalization of the entanglement entropy occurs at the speed of light in consistency with earlier results. This universal regime is the limit where both the length $l$ of the entangling interval and the time $u$ after the quench are taken to infinity with $u/l$ fixed to a value between 0 and $1/2$. This limit captures the exact results for the evolution of the entanglement entropy of entangling intervals with sufficiently large lengths at intermediate times to a very good accuracy. When the entanglement entropy grows linearly, we find that the intersection point of the HRT surface with the null shock is \textit{always behind the horizon of the final black hole} but \textit{always outside the horizon of the initial black hole}. We can solve for the HRT surface analytically in the universal regime.

Secondly, we determine the lower and upper bounds on the irreversible entropy production in quenches leading to transitions between arbitrary thermal states via the quantum null energy condition. We thus generalize the results in \cite{Kibe:2021qjy} which are valid only when the final state is only mildly different from the initial state. We also establish that the transition is always possible when the net change of momentum density vanishes.

\begin{figure}
    \centering
    \includegraphics[scale=0.6]{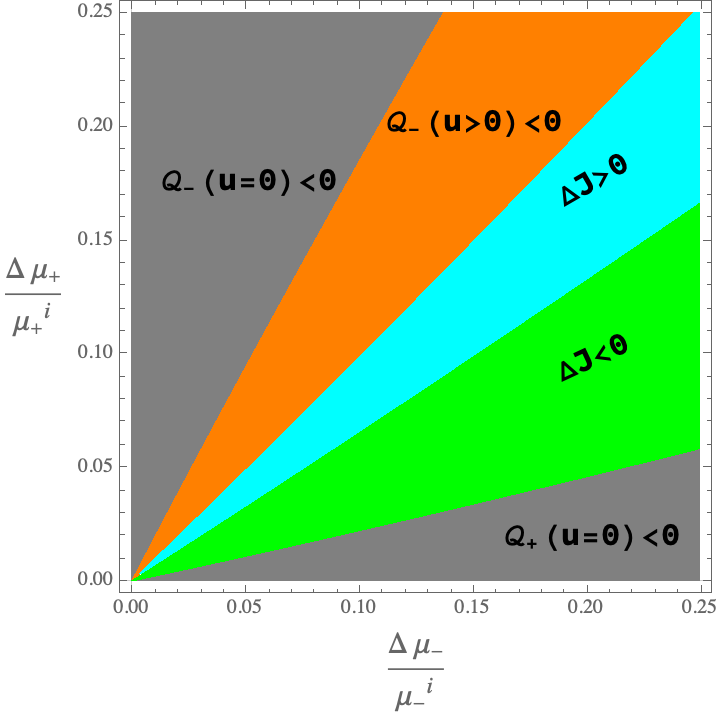}
    \caption{Schematic representation of the allowed final thermal states carrying momenta for a given initial state with fixed temperature and momentum. The allowed final states are in the blue and green regions. See text for details.}
    \label{Fig:schemregplot}
\end{figure}

A schematic representation of our results is in Fig.~\ref{Fig:schemregplot}. In the quenches leading to transitions between thermal states carrying momenta, the non-vanishing components of the energy-momentum tensors of the initial state are $$\langle t_{\pm\pm}\rangle = \frac{c}{12\pi} {\mu_\pm^{i}}^2$$ and those of the final state are $$\langle t_{\pm\pm}\rangle = \frac{c}{12\pi} {\mu_\pm^{f}}^2.$$ Let $\Delta \mu_\pm = \mu_\pm^f - \mu_\pm^i$. Requiring that temperature and entropy should both increase, we obtain  $\Delta \mu_\pm \geq 0$.\footnote{The irreversible entropy production in the quench is simply the difference between initial and final thermodynamic entropies. Note that no reversible entropy is produced due to the heat exchange with the bath as the quench is a unitary process involving an infinite memoryless bath. Furthermore, the increase in both the temperature and entropy follows if we require that the irreversible entropy production should be positive in both the left and right moving sectors of the CFT.} 

Since the initial and final states are quantum equilibrium states, the two QNEC requirements  \eqref{Eq:Qpm-def} are always saturated for any interval of length $l$ before the quench and after time $l/2$ at which the entanglement entropy thermalizes (the HRT surface no longer intersects the shock after the latter time). Without loss of generality, we consider the point $p$ to be the left endpoint of an interval of proper length $l$ so that the displacement of $p$ along the $+$ ($-$) null direction (with the other endpoint held fixed) contracts (expands) the proper length of the interval. Let the quench happen at time $u=0$. With $\mathcal{Q}_\pm$ as defined in \eqref{Eq:Qpm-def}, we will show that
\begin{equation}
    \mathcal{Q}_\pm(u=0) \geq 0 \implies 3 {\mu_\pm^f}^2 - {\mu_\mp^f}^2 \geq 3 {\mu_\pm^i}^2 - {\mu_\mp^i}^2,
\end{equation}
irrespective of the value of $l$. Therefore, QNEC \eqref{Eq:Qpm-def} excludes the gray regions in Fig.~\ref{Fig:schemregplot}. We readily note that the QNEC gives stricter conditions than requiring that the temperature and entropy should both increase as a result of the quench.

It turns out that both the QNEC requirements, with the entanglement evaluated for a semi-infinite interval, are satisfied throughout the quench if they hold at the time of the quench. However, this is not the case for $\mathcal{Q}_- \geq 0$ for a finite length interval, as shown in Fig.~\ref{Fig:schemregplot}. So the negativity of $\mathcal{Q}_-(0<u<l/2)$ for intervals of finite length gives another excluded region which is shown in orange in the figure. The lower boundary of the orange region is actually obtained from the limit $l\rightarrow 0$ in which $\mathcal{Q}_-(u)$ depends on $l$ only through $u/l$ for $0\leq u/l \leq 1/2$, and fixed initial and final states.

Finally, the allowed final states are the blue and green regions of Fig.~\ref{Fig:schemregplot} whose borderline is the curve where the change in total momentum density is zero, i.e. $${\mu_+^f}^2 -{\mu_-^f}^2 = {\mu_+^i}^2 -{\mu_-^i}^2.$$Crucially, the transitions where the total momentum density is unchanged are always allowed. We will discuss in Sec.~\ref{sec:discuss} that this is consistent with the proof of the QNEC in holographic field theories. We note that the bulk energy-momentum tensor localized on the null shock is determined completely by the final and initial states. Generically, whenever the Clausius inequality holds in the dual field theory the classical null energy condition (NEC) is always satisfied by the bulk energy-momentum tensor. However, the QNEC can be violated in the dual field theory even when the energy-momentum tensor on the shock satisfies NEC {(and the bulk also has no naked singularities, c.f. Sec.~\ref{Sec:qnechom-quenches})}. We will discuss the implications of this in Sec.~\ref{sec:discuss}. 

Using thermodynamic identities, we can derive the generalized Clausius inequalities for the allowed final states shown in Fig.~\ref{Fig:schemregplot}. For a given change in the temperature, we obtain a lower and an upper bound for irreversible entropy production.

\subsection{Plan of the paper}
The plan of the paper is as follows. In Sec.~\ref{Sec:qnechom-quenches}, we construct the Ba\~nados-Vaidya metrics describing general quenches leading to transitions between arbitrary quantum equilibrium states. In Sec.~\ref{Sec:qnechom-geodesic-details}, we develop the general algebraic method for the construction of HRT surfaces in these geometries. In Sec.~\ref{sec:qnechom-entropy}, we report general results on the the thermalization of entanglement entropy for quenches between arbitrary thermal states carrying uniform momentum densities. In Sec.~\ref{sec:qnechom-qnecbounds}, we proceed to evaluate whether the QNEC is satisfied in such quenches, determine which final states are allowed for a given initial state, and thus the generalized Clausius inequalities which give lower and upper bounds on irreversible entropy production for a given change in temperature. In Sec.~\ref{sec:recover}, we examine whether the information of the initial state can be recovered from the entanglement entropy and its derivatives, and show that it is only possible at early times. In Sec.~\ref{Sec:QDEC} we discuss the quantum dominant energy conditions. Finally, we conclude with some discussion on the implications of our results in Sec.~\ref{sec:discuss}.


\section{Holographic quenches leading to transitions between quantum equilibrium states}\label{Sec:qnechom-quenches}

\subsection{Ba\~nados-Vaidya metrics}\label{Sec:Prel}
A two-dimensional strongly coupled holographic CFT with a large central charge can be described by three-dimensional Einstein's gravity coupled to a few fields and with a negative cosmological constant $\Lambda$ \cite{Aharony:1999ti}. The gravitational equations of motion take the form
\begin{equation}\label{Eq:Grav}
    R_{MN} - \frac{1}{2} R \,G_{MN} +\Lambda G_{MN} = 8\pi G_N \, T_{MN},
\end{equation}
where $T_{MN}$ is the energy-momentum tensor of the bulk matter fields. Any (time-dependent) state in the CFT corresponds to a solution of the gravitational theory which satisfies regularity conditions and is asymptotically AdS$_3$ with radius $\ell_{AdS}$ given by $\Lambda = - 1/\ell_{AdS}^2$. The central charge of the dual CFT is $c = 3 \ell_{AdS}/(2G_N)$, where $G_N$ is the Newton's gravitational constant \cite{Brown:1986nw,Henningson:1998gx,Balasubramanian:1999re}. Henceforth, we set $\ell_{AdS}=1$.

We will denote the radial coordinate as $z$. The boundary of spacetime where the dual CFT lives is at $z=0$. The boundary is $\mathbb{R}^{1,1}$. We will denote the boundary time and spatial coordinates as $u$ and $y$, respectively. Generic vacuum solutions of Einstein's equations in the ingoing Eddington-Finkelstein coordinates are
\begin{equation}\label{Eq:qnechom-metric}
{\rm d}s^2 = \frac{-2{\rm d}u\,{\rm d}z+(-1+2m(u,y)z^2){\rm d}u^2 + 2j(u,y)z^2 {\rm d}u\,{\rm d}y + {\rm d}y^2}{z^2}
\end{equation}
with
\begin{align}
    m(u,y) = L_+(x^+)+ L_-(x^-), \quad j(u,y) = L_+(x^+)- L_-(x^-)
\end{align}
with $x^\pm = t\pm y$, the future oriented null directions at the boundary. These solutions are the Ba\~nados spacetimes \cite{Banados:1998gg}. Via holographic renormalization \cite{Henningson:1998gx,Balasubramanian:1999re}, we can extract the expectation value of the energy-momentum tensor in the states dual to these solutions. The non-vanishing components of the energy-momentum tensor of these dual states are
\begin{equation}
 \langle t_{\pm\pm}(u,y)\rangle = \frac{c}{12\pi} L_\pm(x^\pm).   
\end{equation}When $L_\pm (x^\pm)$ are constants, these describe Ba\~nados-Teitelboim-Zanelli (BTZ) black holes dual to thermal states with a uniform momentum density. In these cases,
\begin{equation}\label{Eq:Lpm-thermal}
    L_\pm = \mu_\pm^2,
\end{equation}
and the temperature and entropy density are given by
\begin{equation}\label{Eq:TandS}
    T = \frac{2}{\pi} \frac{\mu_+\mu_-}{\mu_+ + \mu_-}, \quad  s = \frac{c}{6}(\mu_+ +\mu_-).
\end{equation}
The generic states dual to Ba\~nados geometries are in the identity conformal block (the universal vacuum sector), and include those which can be obtained via conformal transformations of the vacuum or thermal states. It has been shown in \cite{Ecker:2019ocp} that the quantum null energy condition is saturated in all such states, and following the authors, we will therefore call them \textit{quantum equilibrium states}.

The gravitational solutions describing instantaneous transitions (quenches) between quantum equilibrium states take the form \eqref{Eq:qnechom-metric} (see also \cite{Sfetsos:1994xa}) with 
\begin{align}\label{Eq:m-j-gen}
 m(u,y) &= \theta(-u)(L_+^{i}(x^+) + L_{-}^{i}(x^-)) + \theta(u) (L_+^{f}(x^+) + L_{-}^{f}(x^-)),\\
 j(u,y) &= \theta(-u)(L_+^{i}(x^+) - L_{-}^{i}(x^-)) + \theta(u) (L_+^{f}(x^+) - L_{-}^{f}(x^-)),
\end{align}
where $L_{\pm}^{i,f}(u\pm y)$ are chiral functions that describe the initial, pre-quench ($i$) and final, post-quench ($f$) dual Ba\~nados geometries, respectively. Here, the quench happens at $u=0$. These geometries are supported by a bulk stress tensor $T_{MN}$ that is traceless and locally conserved in the metric \eqref{Eq:qnechom-metric} with non-vanishing components
\begin{equation}\label{Eq:qnechom-Tbulk-gen}
T_{uu} = q(u,y) z+ p(u,y) z^2  + p(u,y) j(u,y) z^3,\qquad T_{uy} = p(u,y)z,
\end{equation}
where $q(u,y)$ and $p(u,y)$ are given by the gravitational constraints. Explicitly,
\begin{align}\label{Eq:qnechom-Tbulk-gen-1}
   & 8\pi G q(u,y) = m(u,y=0)_{\rm dis}= (L_+^{f}(x^+) + L_{-}^{f}(x^-)) - (L_+^{i}(x^+) + L_{-}^{i}(x^-)),\nonumber\\
   & 8\pi G p(u,y) = j(u,y=0)_{\rm dis}= (L_+^{f}(x^+) - L_{-}^{f}(x^-)) - (L_+^{i}(x^+) - L_{-}^{i}(x^-)).
\end{align}
with $m(u,y=0)_{\rm dis}$ and $j(u,y=0)_{\rm dis}$ denoting the discontinuities of $m$ and $j$ at the moment of quench. It is clear that the bulk energy-momentum tensor is localized at the shock and should be thought of as a distribution. It is determined \textit{completely} by the initial and final Ba\~nados geometries, and therefore its form is independent of what matter fields it is constituted of. A natural question therefore is whether such a bulk energy momentum tensor given by \eqref{Eq:qnechom-Tbulk-gen} and \eqref{Eq:qnechom-Tbulk-gen-1} can be actual limits of a smooth solution of Einstein's theory coupled to bulk matter fields whose dynamics are derived from a two-derivative action. As far as we know, this is only true in special cases where the momentum density $j$ vanishes at all times and also the mass density $m$ is homogeneous (i.e. independent of $y$). In this case, such solutions can be regarded as limits of solutions of Einstein's gravity coupled to scalar fields as described in \cite{Bhattacharyya:2009uu}. 

Restricting to transitions between thermal states carrying uniform momentum density, the functions $m$ and $j$ in \eqref{Eq:m-j-gen} reduce to
\begin{align}\label{Eq:m-j-BTZ-transition}
 m(u) &= \theta(-u)({\mu_+^{i}}^{2} + {\mu_-^{i}}^{2}) + \theta(u) ({\mu_+^{f}}^{2} + {\mu_-^{f}}^{2}),\\
 j(u) &= \theta(-u)({\mu_+^{i}}^{2} - {\mu_-^{i}}^{2}) + \theta(u) ({\mu_+^{f}}^{2} - {\mu_-^{f}}^{2})
 \label{Eq:qnechom-BTZ}
\end{align}
by virtue of \eqref{Eq:Lpm-thermal}. In this case, the bulk energy-momentum tensor \eqref{Eq:qnechom-Tbulk-gen-1} takes the following simpler form
\begin{equation}\label{Eq:qnechom-Tbulk}
T_{uu} = q(u) z  + p(u) j(u) z^3,\qquad T_{uy} = p(u)z,
\end{equation}
with
\begin{equation}\label{Eq:qnechom-q-p}
\begin{split}
    8\pi G q(u) =&\ \delta(u)({\mu^f_+}^2 - {\mu^i_+}^2 +{\mu^f_-}^2 - {\mu^i_-}^2), \\ 
    8\pi G p(u) =&\ \delta(u)({\mu^f_+}^2 - {\mu^i_+}^2 -{\mu^f_-}^2 + {\mu^i_-}^2).
\end{split}
\end{equation}

Via holographic renormalization \cite{Henningson:1998gx,Balasubramanian:1999re} we can extract the expectation value of the energy-momentum tensor of the dual state from the metric \eqref{Eq:qnechom-metric} and its non-vanishing components are
\begin{equation}\label{Eq:qnechom-thol-gen}
\langle t_{\pm \pm} \rangle =\frac{c}{12 \pi}\left(\theta(-u) L_\pm^i (x^\pm)  + \theta(u) L_\pm^f(x^\pm) \right),
\end{equation}
which simplifies to
\begin{equation}\label{Eq:qnechom-thol}
\langle t_{\pm \pm} \rangle =\frac{c}{12 \pi}\left(\theta(-u) {\mu_\pm^i}^2  + \theta(u) {\mu_\pm^f}^2 \right)
\end{equation}
for transitions between homogeneous thermal states. It follows that the (homogeneous) energy density $\epsilon$ is given by 
\begin{equation}\label{Eq:qnechom-ehol}
\epsilon = \langle t_{tt} \rangle =\frac{c}{12 \pi}\left(\theta(-u)\left( {\mu_+^i}^2 +  {\mu_-^i}^2\right)   + \theta(u) \left( {\mu_+^f}^2 +  {\mu_-^f}^2\right) \right)
\end{equation}

Gravitational constraints \eqref{Eq:qnechom-Tbulk-gen-1} imply the Ward identity  
\begin{equation}
    \partial_\mu \langle t^{\mu\nu} \rangle = f^\nu,
\end{equation} 
where $f_\nu = (q(u,y), p(u,y))$ is the energy-momentum injection from the infinite bath into the CFT. 

Note that such quenches are unitary processes in the dual CFTs where the Hamiltonian changes instantaneously. Although the explicit form of the change in the Hamiltonian in the dual theory cannot be obtained since the explicit content of the bulk matter has not been specified, we emphasize again that the energy-momentum tensor of the bulk matter (and the energy-momentum injection in the dual theory) are determined via null junction conditions (and Ward identities). 

One can take the point of view that the full geometry \eqref{Eq:qnechom-metric} representing a sharp transition between two Ba\~nados spacetimes across a null shock is an \textit{ensemble averaged} version of the quenches in the full CFT in the large $N$ limit and contain statistical information of the latter (in the same sense that black hole spacetimes contain information of the time-dependent entanglement spectrum of the Hawking radiation \cite{Raju:2020smc,Kibe:2021gtw}\footnote{This should be referred to as the \textit{not-so-fine-grained} entanglement spectrum following \cite{Ghosh:2021axl}. Some averaging is assumed to have been performed in order to treat the Hilbert space as separable. See \cite{Raju:2020smc} for discussions on these issues.}, or solutions of supergravity contain statistical information of heavy operators \cite{Belin:2023efa}). This point of view is natural as the general geometries \eqref{Eq:qnechom-metric} even in the special case of transitions between generic thermal states carrying uniform momentum density are likely to be suitable limits of solutions of classical string theory and not Einstein's gravity coupled to a few scalar fields (we further discuss this in Sec. \ref{sec:discuss}). Nevertheless, as discussed in Sec. \ref{sec:qnechom-entropy}, the time evolution of entanglement entropy after the quench obtained from the holographic prescription is in accordance with what we can expect from the results obtained from the unitary dynamics in  CFTs. The general solutions \eqref{Eq:qnechom-metric} should give quantum statistical information such as the change in the entanglement spectrum and irreversible entropy production in the dual quenches in the large $N$ limit for arbitrary initial and final quantum equilibrium states. 

\subsection{Uniformization map for Ba\~nados spacetimes}\label{Sec:Uniform}


In order to compute entanglement entropy, we will employ uniformization maps which are residual gauge transformations mapping Ba\~nados spacetimes (dual to arbitrary quantum equilibrium states) to Poincar\'{e} AdS$_3$ (dual to the vacuum with vanishing energy-momentum tensor). These bulk diffeomorphisms are uplifts of conformal transformations at the boundary. We use the uniformization maps in the ingoing Eddington-Finkelstein gauge found in \cite{Kibe:2021qjy}.  We reserve uppercase letters $Z,U,Y$ for the Poincar\'{e} AdS$_3$ coordinates, and use lowercase letters $z,u,y$ for Ba\~nados coordinates throughout this paper. Similarly, $X^\pm = T \pm Y$ ($x^\pm = t \pm y$) will denote the boundary null directions for Poincar\'{e} AdS$_3$ (Ba\~nados geometries).

Consider the Poincar\'{e} AdS$_3$ metric 
\begin{equation}
    {\rm d}s^2 = \frac{-2{\rm d}U\,dZ - {\rm d}U^2 + {\rm d}Y^2}{Z^2},
\end{equation}
and the Ba\~nados metric
\begin{equation}
    ds^2 = \frac{{-2{\rm d}z{\rm d}u+(-1+2z^2(L_+(x^+)+L_-(x^-)){\rm d}u^2 + 2z^2(L_+(x^+)-L_-(x^-)){\rm d}u{\rm d}y + {\rm d}y^2}}{z^2}. 
\end{equation}
The uniformization map which takes Ba\~nados coordinates $(z,u,y)$ to Poincar\'{e} coordinates $(Z,U,Y)$ is
\begin{align}\label{eq-uniformization-map}
    Z =& \frac{z\sqrt{{X^+}'(x^+){X^-}'(x^-)}}{1-\frac{z}{2}\lb \frac{{X^+}''(x^+)}{{X^+}'(x^+)} + \frac{{X^-}''(x^-)}{{X^-}'(x^-)} \rb}, \\
    U =& \frac12\lb X^+(x^+) + X^-(x^-) + z\frac{{X^+}'(x^+) + {X^-}'(x^-) - 2\sqrt{{X^+}'(x^+)  {X^-}'(x^-)}}{1-\frac{z}{2}\lb \frac{{X^+}''(x^+)}{{X^+}'(x^+)} + \frac{{X^-}''(x^-)}{{X^-}'(x^-)} \rb} \rb  , \\
    Y =& \frac12 \lb X^+(x^+) - X^-(x^-) + \frac{z\lb {X^+}'(x^+) - {X^-}'(x^-) \rb}{1-\frac{z}{2}\lb \frac{{X^+}''(x^+)}{{X^+}'(x^+)} + \frac{{X^-}''(x^-)}{{X^-}'(x^-)} \rb} \rb , 
\end{align}
where the functions $X^\pm(x^\pm)$ giving the conformal transformation in the dual theory at the boundary (which takes the state dual to the Ba\~nados geometry to the vacuum) are determined via solutions to the equations
\begin{equation}\label{eq-Sch-unif}
    \text{Sch}\lb X^\pm(x^\pm), x^\pm \rb= - 2  L_\pm(x^\pm)
\end{equation}
with Sch denoting the Schwarzian derivative defined as follows:
\begin{equation}
    \Sch\lb f(x), x \rb = \frac{f'''(x)}{f'(x)} - \frac32\frac{f''(x)^2}{f'(x)^2}.
\end{equation}
For general $L_\pm$, these equations are not analytically solvable.

Generally, one can solve \eqref{eq-Sch-unif} via the corresponding linear Hill equation
\begin{equation}
    \Psi''(x) - L(x) \Psi(x) = 0.
\end{equation}
Let the two independent solutions to the Hill equation be $\Psi_1$ and $\Psi_2$. Then, it is easy to check that
\begin{equation}\label{Eq:gen-f}
    f(x) = \frac{a \Psi_1(x)+b \Psi_2(x)}{c \Psi_1(x)+ d \Psi_2(x)}
\end{equation}
(with $a$, $b$, $c$ and $d$ constants) solves
\begin{equation}
    \Sch(f(x),x) = -2L(x).
\end{equation}
So \eqref{Eq:gen-f} with $ad -bc =1$ (to remove the redundancy) corresponds to the general solution of the above equation. The 3-parameter family solutions for $f(x)$ are as expected from the SL$(2,\mathbb{R})$ invariance of the Schwarzian derivative. In the context of the uniformization maps, the general form of $X^\pm(x)$ which solve \eqref{eq-Sch-unif} will have $3 $ parameters each for $X^\pm$, capturing the SL$(2,\mathbb{C})$ isometries of locally AdS$_3$ spacetimes.

In the context of BTZ black holes where $L_\pm = \mu_\pm^2$, we will typically choose
\begin{equation}
    \Psi_1 = \frac1{\sqrt{2\mu}}e^{\mu x},\qquad \Psi_2 = \frac1{\sqrt{2\mu}}e^{-\mu x}
\end{equation}
and $f(x) = \Psi_1/\Psi_2$. We will take advantage of the SL$(2,\mathbb{R})$ transformation of $f(x)$ when needed.

\section{An algebraic procedure for finding HRT surfaces}           \label{Sec:qnechom-geodesic-details}

In a holographic large $N$ strongly interacting 1+1-D CFT, the entanglement entropy $S(A)$ of an interval $A$ in any state is given by \cite{Ryu:2006bv,Hubeny:2007xt},
\begin{equation}\label{Eq:HRT}
    S(A) = \frac{A(\Sigma_A)}{4 G_N}
\end{equation}
where $\Sigma_A$ is the extremal surface (geodesic) in the bulk geometry dual to the state and anchored to the endpoints of $A$ at the boundary. Furthermore, $\Sigma_A$ should be homologous to $A$.\footnote{If there are multiple extremal surfaces satisfying the homology constraint, then one with the minimal area must be chosen. Here we find only unique physical solutions.} The area of $\Sigma_A$ is $A(\Sigma_A)$, which should be regulated by setting a radial cut-off $z=\epsilon$ with $\epsilon$ identified with the short distance cut-off in the CFT. $\Sigma_A$ is is known as the HRT extremal surface.


In this section, we develop an algebraic method for finding the HRT extremal surfaces (geodesics) anchored to endpoints of the entangling interval at the boundary in geometries corresponding to quenches that lead to transitions between quantum equilibrium states, which were discussed in the previous section. We will show that the method allows us to also compute the proper length of the HRT extremal surface and thus evaluate the time-dependence of entanglement entropy algebraically. Our method builds on that developed in \cite{Kibe:2021qjy} but applies even when the final quantum equilibrium state is very different from the initial one. Although our method extends to arbitrary initial and final quantum equilibrium states, here we will focus on transitions between thermal states carrying uniform momentum density, i.e. transitions between arbitrary BTZ black hole spacetimes.\footnote{In absence of quenches, our method reduces simply to the application of uniformization map for computing the holographic entanglement entropy. In fact, uniformization maps have been used earlier to prove the saturation of QNEC in holographic quantum equilibrium states \cite{Ecker:2019ocp}. Uniformization maps have been recently also used to calculate the non-perturbative entanglement entropy in thermal states carrying uniform momentum densities of $T\bar{T}$-deformed holographic CFTs \cite{Banerjee:2024wtl}.}

\subsection{How to cut and glue}          \label{subsec-overview}

\begin{figure}[t]
\centering
\includegraphics[scale=0.55]{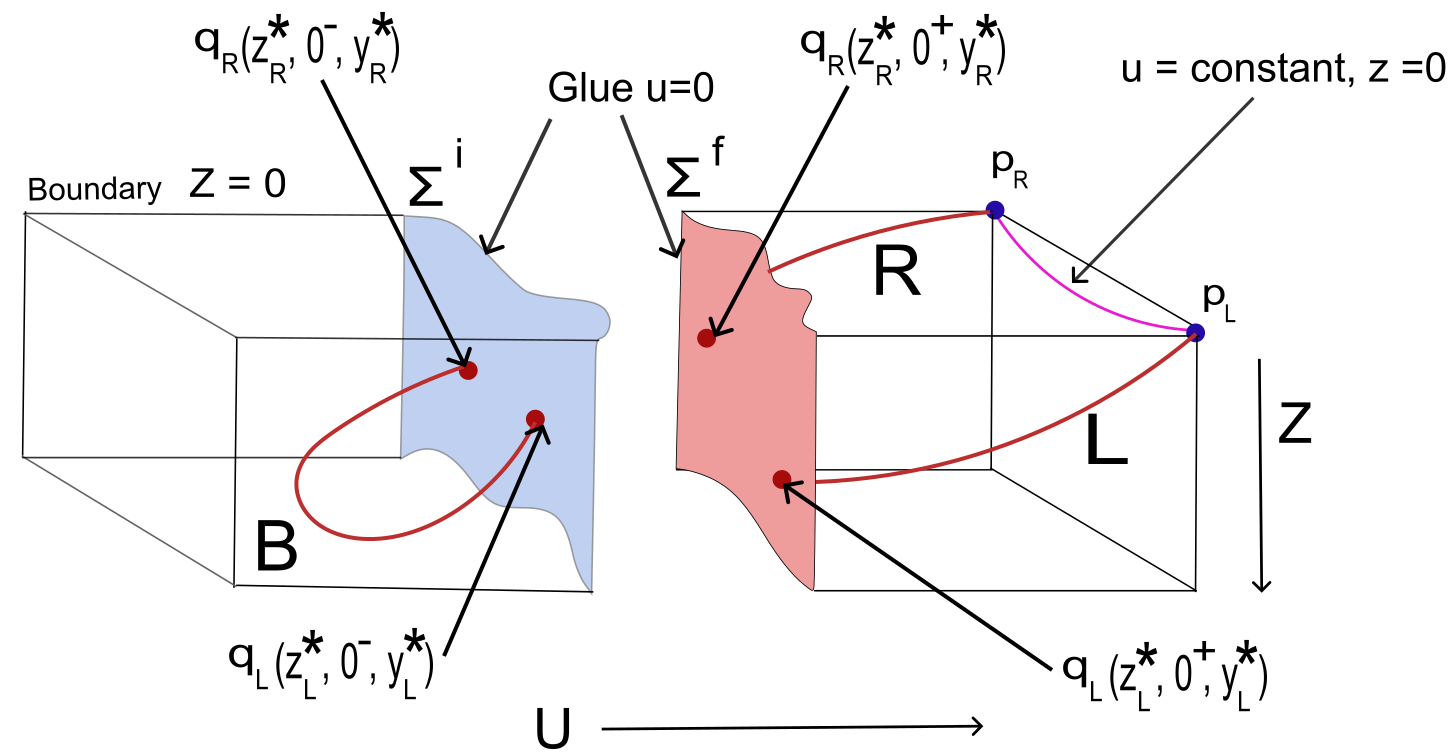}
\caption{Schematic representation of the cut and glue method.  The left and right halves are the initial and final Poincar\'{e} patches. The two hypersurfaces $\Sigma^{i,f}$, which are the images of $u=0$ under the uniformization map in the respective geometries, are shown in blue and red. We glue the two hypersurfaces by identifying points with the same physical coordinate labels. The boundary endpoints of the interval of interest are denoted by $p_{L,R}$. The geodesic ending at $p_{L,R}$ is denoted by a solid red curve. This geodesic is cut into three arcs by the $\Sigma^{i,f}$ hypersurfaces. These arcs are labeled $L,R$ and $B$. The intersection points of the L(R) arc with $\Sigma^f$ are $q_{L(R)}(z^*_{L(R)},u=0^+,y^*_{L(R)})$. These points are identified with the points $q_{L(R)}(z^*_{L(R)},u=0^-,y^*_{L(R)})$ on $\Sigma^i$.}\label{Fig:qnechom-Glue}
\end{figure}

The cut and glue method proposed in \cite{Kibe:2021qjy} essentially makes use of the fact that the geometry \eqref{Eq:qnechom-metric} describing an instantaneous transition between two Ba\~nados spacetimes at $u=0$ can be uniformized, i.e., converted to two Poincar\'{e} patch metrics (with $m(u) = j(u) =0$ in \eqref{Eq:qnechom-metric}) with two \textit{separate} diffeomorphisms, one for $u<0$ and another for $u>0$. These two Poincar\'{e} patches are bounded in the bulk by the hypersurfaces $\Sigma^{i,f} (y,z)$, which are the respective images of the null hypersurface $u=0$ corresponding to the shock under the corresponding uniformization maps. Since the BTZ coordinates are continuous across the shock, the two Poincar\'{e} patches should be glued by identifying points on the hypersurface $\Sigma^i$ with those points on the hypersurface $\Sigma^f$ which have the \textit{same} values of the BTZ coordinates $y$ and $z$. This gluing has been illustrated in Fig.~\ref{Fig:qnechom-Glue}. Since a geodesic segment in Poincar\'{e} AdS$_3$ is part of a semi-circular arc, the uniformization maps allow us to compute the HRT surface (bulk geodesic) algebraically. 

An overview of the key steps of our method to compute the HRT surface anchored to the endpoints of the boundary entangling interval follows. The details are provided throughout the rest of this section. In \cite{Kibe:2021qjy}, the difficulty of solving the extremization condition for the geodesic on the shock surface was avoided by assuming that the final BTZ black hole does not differ significantly from the initial one, i.e. $\mu_\pm^i \sim \mu_\pm^f$. Here we will explicitly solve the extremization conditions at the (two) intersection points where the HRT surface intersects the shock. Thus the method developed here is valid for arbitrary initial and final BTZ black holes.

We first set our notation. We denote the two uniformization maps which map the initial (pre-quench) and final (post-quench) BTZ black hole geometries to Poincar\'{e} patches as $$Z^f(z,u,y),\ U^f(z,u,y),\ Y^f(z,u,y)$$ and $$Z^i(z,u,y),\ U^i(z,u,y),\ Y^i(z,u,y),$$ respectively. We emphasize again that the $z,u,y$ coordinates are continuous. These uniformization maps have been described in Sec.~\ref{Sec:Uniform}. We will denote the boundary ($z\rightarrow 0$) limits of the uniformization maps with a subscript $b$. We recall that $X^{i\pm} = U^i\pm Y^i$ and $X^{f\pm} = U^f\pm Y^f$ reduces to null coordinates $X_b^\pm$ at the boundary, as the boundary restriction of the uniformization map is simply the dual conformal transformation. Note that $X_b^+$ is a function of $x^+$ only, and similarly $X_b^-$ is a function of $x^-$ only. 

As mentioned above, the shock hypersurface $u=0$ has two images
\begin{align}\label{Eq:qnechom-Sigmas}
    \Sigma^{i,f} = (Z^{i,f}(z,0,y),U^{i,f}(z,0,y),Y^{i,f}(z,0,y)). 
\end{align}
under the respective uniformization maps. Points on $\Sigma^{i,f}$ with the same $z$ and $y$ labels are identified. 

For any interval of length $l$ in the dual theory, the entanglement entropy before the quench is simply given by the thermal entanglement entropy,
\begin{equation}\label{Eq:ThermalEE}
    S_{\rm{EE}} = \frac{c}{6}\ln\left(\frac{\sinh(\mu_-^i l)\sinh(\mu_+^i l)}{\mu_-^i\mu_+^i \epsilon^2}\right),
\end{equation}
corresponding to the initial thermal state with $\langle t_{\pm\pm}\rangle$ given by $\mu_\pm^i$ as in \eqref{Eq:qnechom-thol}. The computation of the regularized proper length of the HRT surface (geodesic) easily reproduces this (see below). The short distance cut-off $\epsilon$ appears as the radial cutoff set as $z=\epsilon$ which regularizes the geodetic length in the dual BTZ black hole geometry. Note that before the quench, the HRT surface does not intersect the shock surface (as expected from causality).

We would be interested in computing the entanglement entropy of an arbitrary entangling interval of length $l$ in the dual field theory after the quench. Let the left (right) endpoint $p_{L(R)}$ of the boundary entangling interval be at $x_{L(R)}^\pm$ in the continuous coordinates (and $z_{L(R)} =0$ of course).
Therefore, the Poincar\'{e} patch coordinates of the end-points $p_L$ and $p_R$ are
\begin{equation}\label{Eq:qnechom-Endpoints}
\begin{split}
    & p_L: U_L^f = \frac{1}{2} (X_{b}^{f+}(x_L^+)+X_{b}^{f-}(x_L^-)),\quad  Y_L^f = \frac{1}{2} (X_{b}^{f+}(x_L^+)-X_{b}^{f-}(x_L^-)),\quad Z_L^f =0, \\
    & p_R: U_R^f = \frac{1}{2} (X_{b}^{f+}(x_R^+)+X_{b}^{f-}(x_R^-)),\quad  Y_R^f = \frac{1}{2} (X_{b}^{f+}(x_R^+)-X_{b}^{f-}(x_R^-)),\quad Z_R^f =0.
\end{split}
\end{equation}

As shown in Fig.~\ref{Fig:qnechom-Glue}, the bulk geodesic is cut into three segments when the boundary entangling interval is in the post-quench geometry. We will call the arc connecting the boundary endpoint $p_L$ with the shock surface as $L$, the arc connecting the boundary endpoint $p_R$ with the shock surface as $R$, and the arc lying in the pre-quench geometry as $B$.

We will denote the endpoint of the arc $L$ on the shock surface as $q_L$, and the endpoint of the arc $R$ on the shock surface as $q_R$. Clearly, the arc $B$ connects $q_L$ with $q_R$ in the pre-quench geometry. In the continuous coordinates, $q_L$ is labeled by $y_L^*$  and $z_L^*$, and $q_R$ is labeled by $y_R^*$  and $z_R^*$. We will show below that  $y_{L,R}^*$ and $z_{L,R}^*$ can be obtained via the extremization conditions for the geodesic on the shock surface which can be reduced to algebraic equations. We find that for transitions between two BTZ geometries, the solutions of $y_{L,R}^*$ and $z_{L,R}^*$ are unique. 

The advantage of using uniformization maps is that the computation of the proper lengths of each of the three segments $L$, $R$ and $B$ can be done algebraically also. Once we map the endpoints of each of these segments to the Poincar\'{e} patch via uniformization maps, the length of the geodetic arcs can be computed as follows. If $(Z_1,U_1,Y_1)$ and $(Z_2,U_2,Y_2)$ are the endpoints of a geodesic in the Poincar\'{e} patch, its proper length $L_{\rm geo}$ is given by 
\begin{equation}\label{eq:qnechom-Lgeo}
    L_{\rm geo} = \log(\xi + \sqrt{\xi^2 -1}), \quad \xi = \frac{(Z_1^2 + Z_2^2 - (U_1 +Z_1 - U_2 -Z_2)^2 + (Y_1 - Y_2)^2)}{2 Z_1 Z_2}.
\end{equation}
The entanglement entropy of the boundary interval is then readily obtained via the HRT prescription \eqref{Eq:HRT} which implies that
\begin{equation}
    S_{\rm EE} = \frac{c}{6} \left(L_{\rm geo}^{L}+L_{\rm geo}^{R}+L_{\rm geo}^{B}\right)
\end{equation}
with $L_{\rm geo}^{L}$, $L_{\rm geo}^{R}$ and $L_{\rm geo}^{B}$ denoting the proper lengths of the geodetic segments $L$, $R$ and $B$, respectively (we have used $c= 3/(2G_N)$). The lengths of $L_{\rm geo}^{L}$, $L_{\rm geo}^{R}$ are regulated by considering that the end-points of the entangling interval lie at the regulated boundary $z= \epsilon$. Clearly, $L_{\rm geo}^{L}$, $L_{\rm geo}^{R}$ and $L_{\rm geo}^{B}$ can be readily determined via the corresponding uniformization maps once the continuous coordinate labels  $y_{L,R}^*$ and $z_{L,R}^*$ of the intersection points $q_L$ and $q_R$ between the full codimension two HRT surface (the full geodesic) and the codimension one shock ($u=0$) are determined.

In a nutshell, the algebraic method for determining  $y_{L,R}^*$ and $z_{L,R}^*$ is as follows.
\begin{enumerate}
    \item Use the geodesic equation in the continuous $(z,u,y)$ coordinates to obtain the discontinuities of the $u$ and $y$ components of the tangents to the geodesic (i.e. $\delta\dot{u}$ and $\delta\dot{y}$) on the shock surface. We thus obtain two equations at each of the two intersection points $q_L$ and $q_R$ of the full geodesic with the shock surface.
    \item Express $\dot{u}$ and $\dot{y}$ of the geodetic segments at the two intersection points $q_L$ and $q_R$ in terms of two quantities which are conserved in the \textit{respective} geodetic segments. Note that the two conserved quantities take different values in each geodesic segment.
    \item Assume that the continuous $y$ and $z$ coordinates of the two intersection points $q_L$ and $q_R$ of the geodesics with the shock (at $u=0$) to be $y_{L,R}^*$ and $z_{L,R}^*$. Given the boundary endpoints $p_L$ and $p_R$, evaluate the two conserved quantities in each geodetic segment via the corresponding uniformization map.
    \item Since we know the conserved quantities of each geodesic segment in terms of $y_{L,R}^*$ and $z_{L,R}^*$ via the results of step 3, and the values of $\dot{u}$ and $\dot{y}$ in each geodetic segment at the intersection points in terms of these conserved quantities via the results of step 2, one can readily compute $\dot{u}$ and $\dot{y}$ in the three geodetic segments in terms of $y_{L,R}^*$ and $z_{L,R}^*$ at the intersection points. Finally, use the results of step 1 to obtain four algebraic equations for determining $y_{L,R}^*$ and $z_{L,R}^*$ via $\delta\dot{u}$ and $\delta\dot{y}$ at the two intersection points.
\end{enumerate}
In the rest of this section we elaborate on these steps.

\subsection{Taking the first two steps for obtaining the algebraic equations of the intersection points between HRT surface and the null shock}\label{sec:qnecinhom-btzgeo}


In this subsection, we elaborate on the first two steps for obtaining the intersection points of the co-dimension two HRT surface and the codimension one shock hypersurface. Let us define $f^{i,f}$ and $g^{i,f}$ to be
\[
    f^{i,f} = 2\lb {\mu_+^{i,f}}^2 + {\mu_-^{i,f}}^2 \rb, \qquad g^{i,f} =  {\mu_+^{i,f}}^2 - {\mu_-^{i,f}}^2.
\]

With the spacetime metric given by \eqref{Eq:qnechom-metric} and \eqref{Eq:m-j-BTZ-transition} in case of an instantaneous transition between two arbitrary BTZ black hole geometries, the geodesic equations take the form
\begin{align}\label{Eq:Geo-Eqns}
  &\ddot u + \frac{\dot u^2 - \dot y^2}z = 0, \nonumber\\ 
  &\ddot y  + (g^f-g^i)\dot u^2z^2\delta(u) + \hdots =0,\nonumber\\
  &\ddot z - \frac12z^2\dot u^2 \delta(u)\lb (f^f-f^i) - 2(g^f-g^i)(g^f\theta(u)+g^i\theta(-u))z^2 \rb + \hdots=0,
\end{align}
with the dot denoting differentiation with respect to the proper length parameter. In the above equations we have only retained the terms relevant to compute the discontinuity in $(\dot{z},\dot{u},\dot{y})$, i.e those proportional to $\delta(u)$. The rest of the terms are denoted by dots.  The first of the above equations implies that $\dot u$ is continuous across the null shock at $u=0$, i.e.
\begin{equation}\label{Eq:dotu-jump}
  \delta \dot{u}\vert_{q_{L,R}} = \dot u\vert_{q_{L,R}, u= 0^+} - \dot u\vert_{q_{L,R}, u= 0^-}= 0,
\end{equation}
From the second equation, we obtain that the discontinuity in $\dot{y}$ at the shock is 
\begin{equation}\label{Eq:doty-jump}
  \delta \dot{y}\vert_{q_{L,R}} = \dot y\vert_{q_{L,R}, u= 0^+} -\dot y\vert_{q_{L,R}, u= 0^-} = -(g^f-g^i){z_{L,R}^*}^2\dot u(y_{L,R}^*,z_{L,R}^* ),
\end{equation}
We recall that $y_{L,R}^*$ and $z_{L,R}^*$ are the $y$ and $z$-coordinates of $q_{L,R}$. Note that $\dot{u}$ in the above equation has been evaluated at the intersection point and we have also used that it is continuous at $u=0$. Similarly, from the last equation in \eqref{Eq:Geo-Eqns} we obtain that the discontinuity in $\dot{z}$ at the shock is 
\begin{equation}
	\delta \dot{z}\vert_{q_{L,R}} = \dot z\vert_{q_{L,R}, u= 0^+} -\dot z\vert_{q_{L,R}, u= 0^-} = \frac12 {z^*}^2_{L,R}\dot u\Big( (f^f-f^i) - ({g^f}^2 - {g^i}^2){z^*}^2_{L,R} \Big),
\end{equation}
where we have used $\theta(0)=1/2$. This equation will not be used by us to obtain the algebraic equations for $y_{L,R}^*$ and $z_{L,R}^*$, but will be used to check the accuracy of the solution.

The tangent to the geodesic is $t^\mu=(\dot z,\dot u,\dot y)$. Since the geodesic is spacelike, we should have
\begin{equation}\label{Eq:t-norm}
    t\cdot t = -\frac{2\dot u\dot z+\dot u^2-\dot y^2}{z^2} +f^{i,f}\dot u^2 + 2g^{i,f}\dot u\dot y =1
\end{equation}
at all points of each geodesic segment.

Since $\xi$ and $\chi$ with $\xi\cdot\partial = \partial_u$ and $\chi\cdot\partial = \partial_y$  are Killing vectors in the post-quench and pre-quench geometries, we can obtain the conserved quantities $e_{L}$ and $\lambda_{L}$ for the geodetic segment $L$, $e_{R}$ and $\lambda_{R}$ for the geodetic segment $R$, and $e_{B}$ and $\lambda_{B}$ for the geodetic segment $B$ as follows:
\begin{align}\label{Eq:Conserved-Charges}
  & e_{L} = \xi\cdot t\vert_{q_{L},u=0^+} = \left(f^{f}\dot u + g^{f}\dot y -\frac{\dot z+\dot u}{z^2}\right)\Big\vert_{q_L,u=0^+},   \nonumber\\
  & \lambda_{L} = \chi\cdot t\vert_{q_{L},u=0^+} = \left(\frac{\dot y}{z^2} + g^{f}\dot u\right)\Big\vert_{q_L,u=0^+},\nonumber\\
  & e_{R} = \xi\cdot t\vert_{q_{R},u=0^+} = \left(f^{f}\dot u + g^{f}\dot y -\frac{\dot z+\dot u}{z^2}\right)\Big\vert_{q_R,u=0^+},   \nonumber\\
   & \lambda_{R} = \chi\cdot t\vert_{q_{R},u=0^+} = \left(\frac{\dot y}{z^2} + g^{f}\dot u\right)\Big\vert_{q_R,u=0^+},\nonumber\\
  & e_{B} = \xi\cdot t\vert_{q_{L},u=0^-} = \left(f^{i}\dot u + g^{i}\dot y -\frac{\dot z+\dot u}{z^2}\right)\Big\vert_{q_L,u=0^-},   \nonumber\\
&\quad = \xi\cdot t\vert_{q_{R},u=0^-} = \left(f^{i}\dot u + g^{i}\dot y -\frac{\dot z+\dot u}{z^2}\right)\Big\vert_{q_R,u=0^-},\nonumber\\
  & \lambda_{B} = \chi\cdot t\vert_{q_{L},u=0^-} = \left(\frac{\dot y}{z^2} + g^{i}\dot u\right)\Big\vert_{q_R,u=0^-},\nonumber\\
  &\quad = \chi\cdot t\vert_{q_{R},u=0^-} = \left(\frac{\dot y}{z^2} + g^{i}\dot u\right)\Big\vert_{q_R,u=0^-}
\end{align}
As indicated above, we evaluate the conserved quantities at the intersection points for later convenience. Note that although the conserved quantities take the same values at all points along each geodetic segment, these values depend on the time coordinate(s) of the boundary end points $p_L$ and $p_R$. 

For each geodesic segment $L$, $R$ and $B$, we have three conditions, namely the normalization \eqref{Eq:t-norm} and the expressions of two conserved quantities in \eqref{Eq:Conserved-Charges}. These three conditions can be solved to obtain $(\dot z,\dot u,\dot y)$ at the intersection points in terms of the two conserved charges and $y$ and $z$ coordinates of the intersection points at $u=0$. In case of the segment $L$, we readily find that
\begin{align}\label{Eq:dot-L}
    \dot z(z_L^*, 0^+, y_L^*) =&\ \pm z_L^*\sqrt{1 + (e_L^2-\lambda_L^2-f^f) {z_L^*}^2 + ({g^f}^2 + f^f \lambda_L^2 - 2 e_L g^f \lambda_L ) {z_L^*}^4} \, ,     \nonumber       \\
    \dot u(z_L^*, 0^+, y_L^*) =&\ \frac{-e_L {z_L^*}^2 + g^f \lambda_L {z_L^*}^4 - \dot z(z_L^*, 0^+, y_L^*)}{1 - f^f {z_L^*}^2 + {g^f}^2 {z_L^*}^4}  \, ,   \nonumber  \\
    \dot y(z_L^*, 0^+, y_L^*) =&\ \frac{{z_L^*}^2\lb \lambda_L - (e_L g^f - f^f\lambda_L) {z_L^*}^2 \rb + g^f\dot z(z_L^*, 0^+, y_L^*)}{1 - f^f {z_L^*}^2 + {g^f}^2 {z_L^*}^4}    \, .  
\end{align}
Similarly, for segment $R$, we obtain
\begin{align}\label{Eq:dot-R}
    \dot z(z_R^*, 0^+, y_R^*) =&\ \pm z_R^*\sqrt{1 + (e_R^2-\lambda_R^2-f^f) {z_R^*}^2 + ({g^f}^2 + f^f \lambda_R^2 - 2 e_R g^f \lambda_R ) {z_R^*}^4} \, ,     \nonumber       \\
    \dot u(z_R^*, 0^+, y_R^*) =&\ \frac{-e_R {z_R^*}^2 + g^f \lambda_R {z_R^*}^4 - \dot z(z_R^*, 0^+, y_R^*)}{1 - f^f {z_R^*}^2 + {g^f}^2 {z_R^*}^4}  \, ,   \nonumber  \\
    \dot y(z_R^*, 0^+, y_R^*) =&\ \frac{{z_R^*}^2\lb \lambda_R - (e_R g^f - f^f\lambda_R) {z_R^*}^2 \rb + g^f\dot z(z_R^*, 0^+, y_R^*)}{1 - f^f {z_R^*}^2 + {g^f}^2 {z_R^*}^4}    \, .
\end{align}
and for the segment $B$ we obtain
\begin{align}\label{Eq:dot-B-L}
    \dot z(z_L^*, 0^-, y_L^*) =&\ \pm z_L^*\sqrt{1 + (e_B^2-\lambda_B^2-f^i) {z_L^*}^2 + ({g^i}^2 + f^i \lambda_B^2 - 2 e_B g^i \lambda_B ) {z_L^*}^4} \, ,     \nonumber       \\
    \dot u(z_L^*, 0^-, y_L^*) =&\ \frac{-e_B {z_L^*}^2 + g^i \lambda_B {z_L^*}^4 - \dot z(z_L^*, 0^-, y_L^*)}{1 - f^i {z_L^*}^2 + {g^i}^2 {z_L^*}^4}  \, ,   \nonumber  \\
    \dot y(z_L^*, 0^-, y_L^*) =&\ \frac{{z_L^*}^2\lb \lambda_B - (e_B g^i - f^i\lambda_B) {z_L^*}^2 \rb + g^i\dot z(z_L^*, 0^i, y_L^*)}{1 - f^i {z_L^*}^2 + {g^i}^2 {z_L^*}^4}    \, .  
\end{align}
and
\begin{align}\label{Eq:dot-B-R}
    \dot z(z_R^*, 0^-, y_R^*) =&\ \pm z_R^*\sqrt{1 + (e_B^2-\lambda_B^2-f^i) {z_R^*}^2 + ({g^i}^2 + f^i \lambda_B^2 - 2 e_R g^i \lambda_B ) {z_R^*}^4} \, ,     \nonumber       \\
    \dot u(z_R^*, 0^-, y_R^*) =&\ \frac{-e_B {z_R^*}^2 + g^i \lambda_B {z_R^*}^4 - \dot z(z_R^*, 0^-, y_R^*)}{1 - f^i {z_R^*}^2 + {g^i}^2 {z_R^*}^4}  \, ,   \nonumber  \\
    \dot y(z_R^*, 0^-, y_R^*) =&\ \frac{{z_R^*}^2\lb \lambda_B - (e_B g^i - f^f\lambda_B) {z_R^*}^2 \rb + g^i\dot z(z_R^*, 0^-, y_R^*)}{1 - f^i {z_R^*}^2 + {g^i}^2 {z_R^*}^4}    \, .
\end{align}


This completes the second step for obtaining the algebraic equations for the intersection points. As mentioned before, only the discontinuity in $\dot{u}$ and $\dot{y}$ at the intersection points will be used in practice. For this, we need to choose the correct branch of solution of $\dot{z}$ in \eqref{Eq:dot-L}, \eqref{Eq:dot-R}, \eqref{Eq:dot-B-L} and \eqref{Eq:dot-B-R}. This can be readily done by checking the continuity of $\ddot{z}$ in each geodesic segment (more in the next subsection).

\subsection{The final steps for obtaining the algebraic equations of the intersection points between HRT surface and the null shock}

The next step is to evaluate the two conserved charges of each of the three geodetic segments in terms of $y_{L,R}^*$ and $z_{L,R}^*$ using the uniformization maps. The use of uniformization maps is convenient because the geodesics in the Poincar\'{e} patch take very simple forms as mentioned below. However, we need to deal with an important subtlety. The $z$-coordinates of the intersection points, i.e. $z_{L,R}^*$ can go behind the horizon of the final BTZ black hole geometry if the boundary entangling interval is sufficiently large, although they are always outside the horizon of the initial BTZ black hole geometry for reasons discussed below.\footnote{The authors in \cite{Hubeny:2012ry} point out that extremal surfaces cannot cross horizons in stationary spacetimes. However, for time-dependent scenarios, such as the quench studied here, we know from the results of \cite{Liu:2013iza} (for a transition from vacuum to non-rotating BTZ) that in the presence of shocks, the correct extremal geodesics do indeed fall into the horizon and come back out.} This implies that we need to use an uniformization map such that the final BTZ horizon and the deepest intersection point lies at a finite value of the Poincar\'{e} patch $Z$ coordinate. 

The desired uniformization map can be readily achieved by using the SL(2,$\R$) freedom in choosing $X^\pm(x^\pm)$ in the uniformization map \eqref{eq-uniformization-map}. Instead of choosing


\begin{equation}
    {X^f}^\pm(x) = \frac{1}{2\mu_\pm^f} e^{2\mu_\pm^f x},
\end{equation}
we can choose the SL(2,$\R$) transformed version


\begin{equation}\label{Eq:Xpm-c}
    \tilde {X^f}^\pm = \frac{ {X^f}^\pm }{c_{\rm SL}  {X^f}^\pm + 1},
\end{equation}
with $c_{\rm SL}>0$ and sufficiently large. In this case, we can check that via the uniformization map \eqref{eq-uniformization-map}  the final BTZ horizon at $z^f_h=(\mu_+^f+\mu_-^f)^{-1}$ is now mapped to the finite Poincar\'{e} radial coordinate
\begin{equation}
    Z_h^f = \frac{2\mu_+^f\mu_-^fe^{x^+\mu_+^f+x^-\mu_-^f}}{c_{\rm SL}^2(\mu_+^f + \mu_-^f)e^{2x^+\mu_+^f + 2x^-\mu_-^f} + 2c_{\rm SL}\mu_+^f\mu_-^f(e^{2x^+\mu_+^f} + e^{2x^-\mu_-^f})} \, .
\end{equation}
For generic values of $x^\pm$, it is easy to check that $Z_h^f$ is finite. In practice, we use a large value of $c_{\rm SL}$ by series expanding our expressions around $c_{\rm SL}\to \infty$. The final solutions for $y_{L,R}^*$ and $z_{L,R}^*$ should of course be independent of the choice of $c_{\rm SL}$. 

It is sufficient to do the SL(2, $\R$) transformation of $ {X^f}^\pm$ (and not $ {X^i}^\pm$) because the intersection points do not cross the horizon of the initial BTZ black hole. It is not hard to see why this is the case. Firstly, we note that in a stationary geometry the HRT surface (geodesic) anchored to the endpoints of a boundary interval can never cross the horizon \cite{Hubeny:2012ry}. The geodetic segments $L$ and $R$ have only one end-point at the boundary. If they cross the horizon of the final BTZ black hole, then the other endpoint of these geodetic segments are not at the boundary. The geodetic segment $B$ which lies in the initial BTZ black hole geometry has both endpoints on the shock null hypersurface ($u=0$). However, if there were no shock, then the geodetic segment $B$ would have intersected the boundary. In this case, it would have been a geodesic in a stationary geometry (the initial BTZ black hole), and therefore it should lie entirely outside the initial horizon. This implies that although the intersection points can cross the final BTZ horizon, they should be outside the initial BTZ horizon.

It is important to see explicitly how the geodetic segments cross the horizon of the final BTZ black hole horizon. Actually there are two types of geodesics in the Poincar\'{e} patch in terms of the proper length parameter $\sigma$, which are as follows:
{{
\begin{align}\label{Eq:qnechom-Poincaregeo}
  (Z,U,Y) =&\ \lb \frac{\sech \s}{\sqrt{\Lambda^2-E^2}},\, U_0 - \frac{\sech \s}{\sqrt{\Lambda^2-E^2}} - \frac{E}{\Lambda^2-E^2}\tanh \s,\, Y_0 + \frac{\Lambda}{\Lambda^2-E^2}\tanh \s \rb, \\
  (Z,U,Y) =&\ \lb -\frac{\csch \s}{\sqrt{E^2-\Lambda^2}},\, U_0 + \frac{\csch \s}{\sqrt{E^2-\Lambda^2}} + \frac{E}{E^2-\Lambda^2}\coth \s,\, Y_0 - \frac{\Lambda}{E^2-\Lambda^2}\coth \s\rb.
\end{align}}}
The first solution is valid for $\Lambda>E$, and the second one for $E>\Lambda$, where $E$ and $\Lambda$ are two integration constants (and also conserved quantities). Also $U_0$ and $Y_0$ are integration constants. In the first branch, $\sigma\in (-\infty, \infty)$ and the geodesic reaches the boundary $Z=0$ for $\sigma = \pm \infty$. The boundary endpoints (at $Z=0$) are:
{\small{
\begin{equation}\label{Eq:end-1-2}
    p_1: U =U_0 + \frac{E}{\Lambda^2-E^2},\, Y= Y_0 - \frac{\Lambda}{\Lambda^2-E^2} ,\quad p_2: U = U_0 - \frac{E}{\Lambda^2-E^2},\,  Y= Y_0 + \frac{\Lambda}{\Lambda^2-E^2}.
\end{equation}}}
In the second branch, $\sigma \in (-\infty,0)$ (since $Z\geq 0$) -- the geodesic starts from the boundary and continues to the Poincar\'{e} horizon ($Z=\infty$) without coming back to the boundary. The boundary endpoint is at
\begin{equation}\label{Eq:end-1}
    p: U = U_0 - \frac{E}{\Lambda^2-E^2}, \,Y= Y_0 + \frac{\Lambda}{\Lambda^2-E^2}.
\end{equation}

Nevertheless, it turns out we do not need to know explicitly which of the two branches of the geodesics on the Poincar\'{e} patch is obtained when we apply the uniformization maps to the geodetic segments $L$ and $R$. To see this, consider a geodesic in the Poincar\'{e} patch connecting the points $(Z_1, U_1, Y_1)$ and $(Z_2, U_2, Y_2)$. We can solve for the four integration constants $E$, $\Lambda$, $U_0$ and $Y_0$ along with the two values of $\sigma$, namely $\sigma_1$ and $\sigma_2$ at which the endpoints are attained in terms of the endpoint coordinates $(Z_1, U_1, Y_1)$ and $(Z_2, U_2, Y_2)$. Remarkably, in both of these branches we obtain the same forms of $E$ and $\Lambda$ given in terms of $(Z_1, U_1, Y_1)$ and $(Z_2, U_2, Y_2)$ which explicitly are
{\small{
\begin{align}\label{eq-epslam-sol}
    \Lambda =&\ \frac{2(Y_2-Y_1)}{\sqrt{\big(-(Y_2-Y_1)^2+(U_1-U_2+2Z_1)(U_1-U_2-2Z_2)\big)\big( -(Y_2-Y_1)^2+(U_1-U_2)(U_1+2Z_1-U_2-2Z_2) \big)}}, \nn  \\
    E =&\ \frac{U_1+Z_1-U_2-Z_2}{Y_2-Y_1}\,\Lambda\,.       
\end{align}}}
Clearly from \eqref{Eq:qnechom-Poincaregeo} we readily see that the values of $\sigma$ at the endpoints can be obtained from the $Z$ coordinates of the endpoints. Explicitly, in the first branch
\begin{equation}\label{Eq:s1-s2}
  \sigma_1 = -{\rm arccosh} \frac{1}{Z_1\sqrt{\Lambda^2-E^2}}, \quad \sigma_2 = {\rm arccosh} \frac{1}{Z_2\sqrt{\Lambda^2-E^2}}
\end{equation}
with $E$ and $\Lambda$ as in \eqref{eq-epslam-sol}. Finally, we can extract $U_0$ and $Y_0$ from the $U$ and $Y$ coordinates of any of the two endpoints. Explicitly, in the first branch
\begin{equation}\label{Eq:U0-Y0}
    U_0 = U_1 + \frac{\sech \sigma_1}{\sqrt{\Lambda^2-E^2}} + \frac{E}{\Lambda^2-E^2}\tanh \sigma_1, \quad Y_0 = Y_1 - \frac{\Lambda}{\Lambda^2-E^2}\tanh \sigma_1.
\end{equation}
with $E$, $\Lambda$, $\sigma_1$ and $\sigma_2$ as in \eqref{eq-epslam-sol} and \eqref{Eq:s1-s2}. We can similarly proceed with the other branch. We thus obtain $E$, $\Lambda$, $U_0$ and $Y_0$ from the $(Z,U,Y)$ coordinates of the endpoints (or any two points) of the geodesic. If one of the endpoints is at the boundary ($Z=0$), then one can use a simpler procedure to extract $U_0$ and $Y_0$ via \eqref{Eq:end-1-2} or \eqref{Eq:end-1} (depending on the branch) and \eqref{eq-epslam-sol}.

With the details presented above, we can readily state the third step of finding the algebraic expressions of the conserved charges $e_{L,R,B}$ and $\lambda_{L,R,B}$ of the respective geodetic segments in terms of $y_{L,R}^*$ and $z_{L,R}^*$. Let us assume that the $y$ and $z$ coordinates of $q_{L,R}$ are $y_{L,R}^*$ and $z_{L,R}^*$, respectively. Given the coordinates of the endpoints $p_L$, $p_R$, $q_L$, $q_R$ in the $(z,u,y)$ coordinates, we can use the uniformization map to obtain the Poincar\'{e} patch coordinates of all these points. Thus we know the Poincar\'{e} patch coordinates of the endpoints of all the geodetic segments $L$, $R$ and $B$. Using \eqref{eq-epslam-sol}, we can obtain the parameters $E$ and $\Lambda$ of the respective geodesic (now mapped to the Poincar\'{e} patch), and then the other parameters $U_0$ and $Y_0$, along with the proper length parameters $\sigma_1$ and $\sigma_2$ labeling the endpoints, using the procedure mentioned above. For the latter we need to assume that the geodesic in the Poincar\'{e} patch belongs to one of the two branches. It turns out that if we use the uniformization map with ${X^f}^\pm$ given by \eqref{Eq:Xpm-c}, then we can find solutions for $\sigma_1$, $\sigma_2$, $U_0$ and $Y_0$ using only the first branch (if $c_{\rm SL}$ is sufficiently large) for each geodetic segment. Thus we can algebraically find out all the parameters of each geodetic segment mapped to the Poincar\'{e} patch. 

We can then readily proceed to obtain the algebraic forms of the conserved charges $e_{L,R,B}$ and $\lambda_{L,R,B}$ of the respective geodetic segments in terms of $y_{L,R}^*$ and $z_{L,R}^*$. As for instance, consider the $L$ segment mapped to the Poincar\'{e} patch. We can evaluate $e_L$ via
\begin{equation}
    e_L = t\cdot \xi \,\, {\rm with}\,\, t^\mu = \lb \dot{Z},\dot{U},\dot{Y} \rb\vert_{\sigma =\sigma_1}, \,\, \xi^\mu = (\partial_u Z^f, \partial_u U^f, \partial_u Y^f)\vert_{q_L}
\end{equation}
and $\lambda_L$ via
\begin{equation}
    \lambda_L = t\cdot \chi \,\, {\rm with}\,\, t^\mu = \lb \dot{Z},\dot{U},\dot{Y} \rb\vert_{\sigma =\sigma_1}, \,\, \xi^\mu = (\partial_y Z^f, \partial_y U^f, \partial_y Y^f)\vert_{q_L}
\end{equation}
where we should use the uniformization map given by \eqref{eq-uniformization-map} and \eqref{Eq:Xpm-c} to obtain the Poincar\'{e} patch coordinates of $q_L$ in terms of $y_L^*$ and $z_L^*$ of $q_L$ (at which the proper length parameter of the Poincar\'{e} patch geodesic takes the value $\sigma_1$) and use the Poincar\'{e} patch metric to compute the inner product. Note that via the uniformization map we also obtain the components of $\xi$ and $\chi$ explicitly in the Poincar\'{e} patch also  in terms of $y_L^*$ and $z_L^*$. Thus we obtain $e_L$ and $\lambda_L$ in terms of $y_L^*$ and $z_L^*$. Similarly we can proceed to compute the conserved charges of the other geodetic segments. The expressions are rather long and therefore we do not quote them here. This concludes the third step for obtaining the algebraic equations for $y_{L,R}^*$ and $z_{L,R}^*$.

The final step for obtaining the four algebraic equations for determining $y_{L,R}^*$ and $z_{L,R}^*$ is straightforward. These four equations are obtained by computing the jump in $\dot{u}$ and $\dot{y}$ at the intersection points $q_L$ and $q_R$ in two different ways, and equating them with each other. In both ways the result is given in terms of $y_{L,R}^*$ and $z_{L,R}^*$ (note that the boundary endpoints of the HRT surface, $p_L$ and $p_R$ should be already specified). In the first way, we compute $\delta\dot{u}$ and $\delta\dot{y}$ by integrating the geodesic equation in the continuous coordinates across the shock resulting in \eqref{Eq:dotu-jump} and \eqref{Eq:doty-jump}. In \eqref{Eq:doty-jump}, we can specify 
$\dot{u}$ at $q_{L}$ (or $q_R$) using \eqref{Eq:dot-B-L} (or \eqref{Eq:dot-B-R}) obtained in the second step and the expressions for the conserved charges of the $B$ segment in terms of $y_{L,R}^*$ and $z_{L,R}^*$ obtained in the third step. The second way to compute the jump in $\dot{u}$ and $\dot{y}$ at $q_L$ and $q_R$ is by using the expressions for $\dot{u}$ and $\dot{y}$ before the shock in \eqref{Eq:dot-B-L} (or \eqref{Eq:dot-B-R}, and after the shock in \eqref{Eq:dot-L} (or \eqref{Eq:dot-R}). Furthermore, we also use the explicit form of the conserved charges $e_{L,R,B}$ and $\lambda_{L,R,B}$ of the respective geodetic segments given in terms of $y_{L,R}^*$ and $z_{L,R}^*$ obtained in the third step. We find that there are unique real solutions for $y_{L,R}^*$ and $z_{L,R}^*$ with $z_{L,R}^* \geq 0$. 

In most cases, the four algebraic equations for $y_{L,R}^*$ and $z_{L,R}^*$ can be solved only numerically. However, in some physically interesting limits, we can solve the equations analytically as mentioned in the following sections.

With the assumption, as in our earlier work \cite{Kibe:2021qjy}, that
\begin{equation}\label{eq:qnechom-perturbative}
    l(\mu_\pm^f-\mu_\pm^i) \ll 1,
\end{equation}
where $l$ is the length of the boundary interval, both $\dot{u}$ and $\dot{y}$ of the geodesic are continuous at the shock at leading order. In this case, one can obtain the intersection points by first obtaining the geodesic, with endpoints $p_{L,R}$, in the final BTZ geometry, assuming there is no shockwave. One then obtains the intersection points by simply solving for the intersection of this geodesic with the shock hypersurface ($u=0$). Employing the uniformization maps, we can convert the latter to purely algebraic equations. When \eqref{eq:qnechom-perturbative} is valid, one expects only a small difference between the geodesics with and without a shockwave. We will also see that although we get different answers for $y_{L,R}^*$ and $z_{L,R}^*$ when the assumption \eqref{eq:qnechom-perturbative} is invalid, we get exactly same results as those obtained assuming that both $\dot{u}$ and $\dot{y}$ of the geodesic are continuous at the shock, for many physically interesting quantities in some limits.

\section{Evolution and thermalization of entanglement entropy}\label{sec:qnechom-entropy}

We first study the time-evolution of entanglement entropy of an interval during quenches between thermal states carrying uniform momentum density. Due to spatial translation invariance, we can set the left end point of the interval $p_L$ at $(0,u,0)$ and the right end point of the interval $p_R$ at $(0,u,l)$ in the continuous $(z,u,y)$ coordinates, without loss of generality. The length of the interval is $l$. We use the algebraic method presented in the previous section to calculate the evolution of the entanglement entropy of the interval. In general, the algebraic equations for the intersection points between the HRT surface and the shock hypersurface need to be solved numerically, however we obtain general and exact analytic results in certain limits illuminating each stage of the evolution of entanglement entropy as described in this section.

Let us proceed with the assumption that the classical Clausius inequality (which is nothing but the bulk NEC) is satisfied by the quench for both left and right moving sectors, i.e. $\mu_\pm^f \geq \mu_\pm^i$. In the following section, we will derive the generalized Clausius inequalities by studying the QNEC. Prior to the quench, i.e. for $u< 0$, the entanglement entropy of the interval is of the thermal form as given by \eqref{Eq:ThermalEE} with $\mu_\pm$ set to initial values.

The time evolution of entanglement entropy of the interval of length $l$, which is a continuous function of time, has three distinct phases:
\begin{enumerate}
    \item \textbf{Early time quadratic growth}: Just after the quench, the entanglement entropy of the interval grows as $u^2$ with a coefficient which is determined only by the change in the (uniform) energy density due to the quench and is independent of $l$, the length of the interval for arbitrary $\mu_\pm^{i,f}$. 
        \item \textbf{Intermediate time linear growth}: At intermediate times, the entanglement entropy grows linearly. In the double scaling limit where $u\rightarrow\infty$ and $l\rightarrow\infty$ with $u/l$ fixed to any value between $0$ and $1/2$, the coefficient of the linear growth can be exactly determined for arbitrary $\mu_\pm^{i,f}$ only in terms of the change in the (uniform) entropy density.
    \item \textbf{Approach to equilibrium}: The entanglement entropy attains the final equilibrium value exactly at $u=l/2$. For times $u\approx\frac{l}{2}$, the difference between the final value of the entanglement entropy and that at time $u$ behaves as $\left(\frac{l}{2}-u\right)^{3/2}$. Both of these are valid for arbitrary $\mu_\pm^{i,f}$.
\end{enumerate}

\begin{figure}
\centering     
\subfigure[]{\includegraphics[scale=0.78]{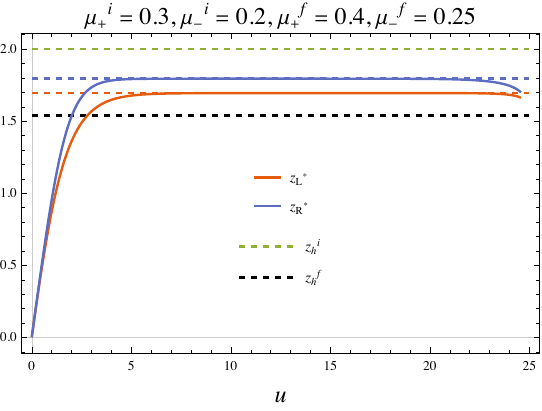}}~~
\subfigure[]{\includegraphics[scale=0.78]{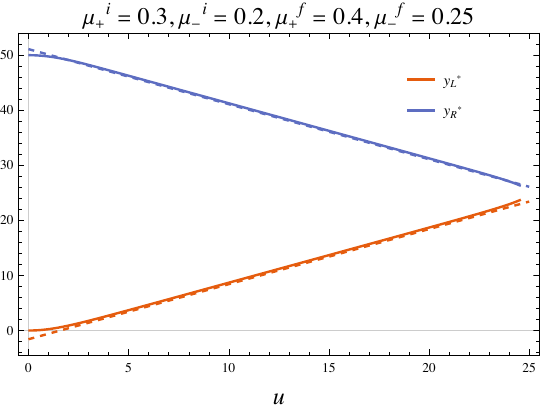}}
\caption{Plots of the evolution of the intersection points $z_{L,R}^*$  (left) and $y_{L,R}^*$ (right) for a boundary interval of length $l=50$ satisfying $\mu_\pm^{i,f}l \gg 1$.  We have chosen $\mu_+^i = 0.3$, $\mu_-^i =0.2$, $\mu_+^f = 0.4$ and $\mu_-^f =0.25$. The dashed red (and blue) curves indicate the analytic expressions for $z_{L}^*$ (and $z_R^*$) on left and $y_{L}^*$ (and $y_R^*$) on right in the double scaling limit \eqref{Eq:double-scale}. These are explicitly given by \eqref{Eq:double-scale-zLR-yLR}. The $z$ coordinates of the initial and final horizons $z_h^{i,f}$ are indicated with dashed lines on the left. We readily see that the intersection points cross the final but not the initial horizon. On the right we see that in the intermediate regime $y_L$ and $y_R$ approach each other at speed of light at intermediate times converging eventually at $l/2$ at time $u = l/2$ (note $l/2 = 25$) as indicated in \eqref{Eq:z-y-converge}. At $u=l/2$, similarly $z_{L}^*$ and $z_R^*$ converge to the value given by \eqref{Eq:z-y-converge}. }
\label{Fig:qnechom-intpts}
\end{figure}

The key to obtaining these results is solving for the (continuous) $y$ and $z$ coordinates of the intersection points of the HRT surface with the null shock hypersurface $u=0$. A representative plot for the evolution of the $z_{L,R}^*$ coordinates of the left and right intersection points $q_L$ and $q_R$ is shown in Fig. \ref{Fig:qnechom-intpts} for a sufficient large $l$ (i.e. for $\mu_\pm^{i,f}l \gg 1$). As expected, both $z_{L,R}^*$ move from the boundary to the interior for $u\geq 0$. At early time, $z_{L,R}^*$ can be solved in a series expansion in $u$. We readily note that at intermediate times, $z_{L,R}^*$ stay constant at a value which is behind the final black hole horizon but outside the initial black hole horizon. For $\mu_\pm^{i,f}l \gg 1$, the values where $z_{L,R}^*$ almost plateau (saturate) at intermediate times coincide very well with the analytic solutions of $z_{L,R}^*$ obtained in the double scaling limit $u\rightarrow\infty$ and $l\rightarrow\infty$ with $u/l$ fixed to any value between $0$ and $1/2$ (c.f. Fig. \ref{Fig:qnechom-intpts}). In case of transition from vacuum to a thermal state carrying no momentum, the saturation values of $z_{L,R}^*$ is twice the $z$-coordinate of the final black hole horizon as shown earlier in \cite{Liu:2013iza,Liu:2013qca}. However, the penetration of $z_{L,R}^*$ behind the final black hole horizon is less when the initial and final black holes carry momentum.

The HRT surface does not intersect the shock for $u>l/2$ and therefore the entanglement entropy thermalizes at $u=l/2$. The final values of $z_{L,R}$ at $u=l/2$ can be obtained analytically. Note that $z_L^*$ and $z_R^*$ should coincide at $u=l/2$ as the two intersection points $q_L$ and $q_R$ coincide to a single point which is the last intersection point between the shock and the HRT surface. The final value of $z_{L,R}^*$ coincide with the final black hole horizon when $\mu_\pm^fl$ is large,

Similarly, $y_{L}^*$ and $y_R^*$ coincide with the $y$-coordinates of the endpoints of the boundary interval, $p_L$ and $p_R$, respectively at $u=0$. As shown in Fig. \ref{Fig:qnechom-intpts}, subsequently they move close together. At early time, the values of $y_{L,R}^*$ can be solved perturbatively. At intermediate times, $y_{L,R}^*$ approach each other at the speed of light. In this regime, for $\mu_\pm^{i,f}l \gg 1$, $y_{L,R}^*$ also match with the analytic values obtained by solving the algebraic equations for $z_{L,R}^*$ and $y_{L,R}^*$ in the double scaling limit $u\rightarrow\infty$ and $l\rightarrow\infty$ with $u/l$ fixed to any value between $0$ and $1/2$ (c.f. Fig. \ref{Fig:qnechom-intpts}). The final values of $y_{L}$ and $y_R$ are $l/2$, i.e. they coincide with the $y$-coordinate of the mid-point of the boundary interval (as $q_L$ and $q_R$ coincide).

In what follows, we will provide the details of the above results in each of the three regimes. It is worth emphasising that we can always numerically solve for $y_{L,R}$ and $z_{L,R}$. Note that we need sufficiently high numerical precision to accurately find the roots of the algebraic equations. This is because of exponential terms in the equations, such as $e^{-\mu_\pm^{i,f}l}$ and $e^{\mu_\pm^{i,f}l}$, and similar terms involving the time $u$.

\subsection{Early time quadratic growth regime}
For arbitrary $\mu_\pm^{i,f}$ and $l$, the solutions for the $y$ and $z$ coordinates of the intersection points between the HRT surface and the shock at small and positive $u$ are 
\begin{align}
    z_L^* &= u + \frac{1}{2}\left(\mu_{-}^{i} \coth(\mu_{-}^{i} l)-\mu_{+}^{i} \coth(\mu_{+}^{i} l) \right) u^2 +\mathcal{O}(u^3),\nonumber\\
    y_L^* &= \frac{1}{2}\left(\mu_{-}^{i} \coth(\mu_{-}^{i} l)+\mu_{+}^{i} \coth(\mu_{+}^{i} l) \right) u^2+\mathcal{O}(u^3),\nonumber \\
    z_R^* &= u - \frac{1}{2}\left(\mu_{-}^{i} \coth(\mu_{-}^{i} l)-\mu_{+}^{i} \coth(\mu_{+}^{i} l) \right) u^2+\mathcal{O}(u^3),\nonumber\\
    y_R^* &= l - \frac{1}{2}\left(\mu_{-}^{i} \coth(\mu_{-}^{i} l)+\mu_{+}^{i} \coth(\mu_{+}^{i} l) \right) u^2+\mathcal{O}(u^3).
\end{align}
From these solutions, we obtain that the entanglement entropy behaves as
\begin{equation}\label{Eq:earlytimeEE}
    S_{\rm EE} = S_{\rm EE}^{\rm in} + \frac{c}{6} \left({\mu_{+}^{f}}^2+{\mu_{-}^{f}}^2- {\mu_{+}^{i}}^2-{\mu_{-}^{i}}^2\right) u^2 +\mathcal{O}(u^3).
\end{equation}
where $S_{\rm EE}^{\rm in} $ denotes the initial thermal entropy before the shock as given by \eqref{Eq:ThermalEE} for an arbitrary entangling lengh $l$. The above implies that the rate of quadratic growth of entanglement entropy at early time for arbitrary entangling length $l$ is
\begin{equation}\label{Eq:D}
    \mathcal{D} =  \frac{c}{6} \left({\mu_{+}^{f}}^2+{\mu_{-}^{f}}^2- {\mu_{+}^{i}}^2-{\mu_{-}^{i}}^2\right) = 2\pi \Delta\varepsilon,
\end{equation}
where $ \Delta\varepsilon$ is the change in the energy density as the result of the quench (c.f. Eq. \eqref{Eq:qnechom-ehol}). Thus, at early times the rate of growth of entanglement is determined simply by the change in energy density in an arbitrary quench and is independent of the entangling length $l$. Our result for $\mathcal{D}$ agrees with that in \cite{Liu:2013iza,Liu:2013qca,Hubeny:2013hz,Ageev:2017wet} in the limit when both the initial and final spacetimes are non-rotating BTZ black branes. Our result was perturbatively established for transitions between rotating BTZ spacetimes in \cite{Kibe:2021qjy} assuming small $\mu_\pm^f-\mu_\pm^i$. The early-time quadratic growth of entanglement entropy has also been seen in numerical simulations of lattice systems \cite{Unanyan:2010,Unanyan:2014}.

\begin{figure}
\centering     
\subfigure[]{\includegraphics[scale=0.78]{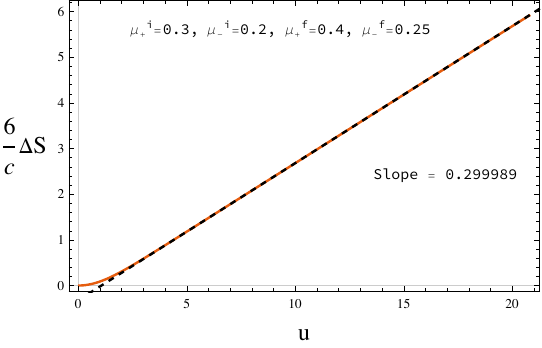}}
\subfigure[]{\includegraphics[scale=0.78]{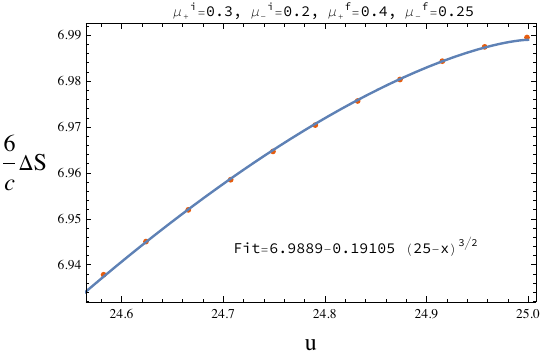}}
\caption{Here $\Delta S(u) = S_{\rm EE}(u) -  S_{\rm EE}^{\rm in}$, the growth of entanglement entropy from its initial value $ S_{\rm EE}^{\rm in}$ for an entangling interval of length $50$ after the quench at $u=0$. We have chosen $\mu_+^i = 0.3$, $\mu_-^i =0.2$, $\mu_+^f = 0.4$ and $\mu_-^f =0.25$ for both plots. Note that thermalization of the entanglement entropy happens at $u=l/2 = 25$. (a) The intermediate regime for $(6/c)\Delta S$ fits very well with a linear function having slope $(12/c)(s^f-s^i)= 2(\mu_+^f +\mu_-^f -\mu_+^i -\mu_-^i) = 0.3$ as implied by \eqref{Eq:qnechomo-vs}. (b) The late time regime $u \approx \frac{l}{2}$ fits well with $\frac32$ saturation exponent as implied by \eqref{Eq:Exp-Sat}.}
\label{Fig:qnechom-entropy}
\end{figure}

\subsection{Linear growth regime and the double scaling (steady state) limit}
\label{Sec:lingrowth}
The intermediate regime arises for large entangling intervals for which $\mu_\pm^{i,f}l\gg 1$. As discussed above, we can analytically access this regime via considering the double scaling limit where 
\begin{equation}\label{Eq:double-scale}
    l \to \infty, \quad u \to \infty, \quad 0<\frac{u}{l} \leq \frac{1}{2},
\end{equation}
In this limit, we obtain the following exact solutions for the $y$ and $z$ coordinates of $q_L$ and $q_R$, the intersection points of the HRT surface and the shock. Explicitly, these solutions for arbitrary $\mu_\pm^{i,f}$ are
\begin{align}\label{Eq:double-scale-zLR-yLR}
    z^*_L &=\frac{2}{\mu_-^f+ \mu_+^i+\sqrt{2({\mu_-^i}^2+{\mu_+^f}^2) -(\mu_-^f-\mu_+^i)^2}}\nonumber\\
    y^*_L &=u +\frac{1}{2 \mu_+^f} \log \left[\frac{\mu_-^i- \mu_+^f+\sqrt{2({\mu_-^i}^2+{\mu_+^f}^2) -(\mu_-^f-\mu_+^i)^2}}{\mu_-^i+ \mu_+^f+\sqrt{2({\mu_-^i}^2+{\mu_+^f}^2) -(\mu_-^f-\mu_+^i)^2}} \right]\nonumber\\
    z^*_R &=\frac{2}{\mu_+^f+ \mu_-^i+\sqrt{2({\mu_+^i}^2+{\mu_-^f}^2) -(\mu_+^f-\mu_-^i)^2}}\nonumber\\
    y^*_R&=l -u - \frac{1}{2 \mu_-^f} \log \left[\frac{\mu_+^i- \mu_-^f+\sqrt{2({\mu_+^i}^2+{\mu_-^f}^2) -(\mu_+^f-\mu_-^i)^2}}{\mu_+^i+ \mu_-^f+\sqrt{2({\mu_+^i}^2+{\mu_-^f}^2) -(\mu_+^f-\mu_-^i)^2}} \right],
\end{align}
assuming $\mu_+^f>\mu_-^f$. (Otherwise the expressions for $z_{L,R}^*$ are exchanged and the expression for $y_L^* -u$ is exchanged with that for $l-u- y_R^*$.) Note that such an assumption is needed to decide an order relation between terms such as $e^{\mu^f_\pm l}$ which appear in the large $l$ analysis.
Clearly the $z$ coordinates of the intersection points are constants while the $y$ coordinates of the intersection points approach each other at the speed of light. As shown in Fig.~\ref{Fig:qnechom-intpts} and discussed above, these analytic solutions match very well with the numerical solutions when $\mu_\pm^{i,f}l\gg 1$ at intermediate times.

We can readily note from \eqref{Eq:double-scale-zLR-yLR} that $z_{L,R}^*$ satisfy
\begin{equation}
\frac{1}{\mu^f_++\mu_-^f} \leq z_{L,R}^* \leq \frac{1}{\mu^i_++\mu_-^i}. 
\end{equation}
given that $\mu_\pm^f \geq \mu_\pm^i$ (which we have assumed from the beginning). The lower bound above is the $z$ coordinate of the outer horizon of the final BTZ black hole while the upper bound is the  $z$ coordinate of the outer horizon of the initial BTZ black hole. Thus the intersection points are behind the outer horizon of the final black hole and outside the outer horizon of the initial black hole. 

It follows from \eqref{Eq:double-scale-zLR-yLR} that in this limit the entanglement entropy grows linearly with time with a slope given by
\begin{equation}\label{Eq:qnechomo-vs}
   \frac{{\rm d} S_{\rm EE}}{{\rm d}u}  = \frac{c}{3}\lb \mu_+^f + \mu_-^f - \mu_+^i- \mu_-^i \rb = 2\times \left(s^f-s^i\right).
\end{equation}
For the second equality above we have used \eqref{Eq:TandS}. To interpret this physically, we note from \eqref{Eq:ThermalEE} that for large lengths satisfying $\mu_\pm^{i,f}l \gg 1$, the initial and final entanglement entropies are simply $s^i l$ and $s^f l$, respectively. Therefore, \eqref{Eq:qnechomo-vs} implies that the entanglement entropy thermalizes at the speed of light from both ends of the interval in the double scaling limit \eqref{Eq:double-scale}. This feature is known as the \textit{entanglement tsunami} phenomenon \cite{Liu:2013iza}. As shown in the representative numerical plot in Fig.~\ref{Fig:qnechom-intpts} (a), the linear growth of entanglement entropy with slope given by \eqref{Eq:qnechomo-vs} is reproduced very well for $\mu_\pm^{i,f}l \gg 1$ at intermediate time.

This result was shown analytically for transitions between non-rotating BTZ spacetimes \cite{Liu:2013iza,Liu:2013qca,Hubeny:2013hz,Ageev:2017wet}. In \cite{Kibe:2021qjy}, this result was also obtained in the approximation that the initial and final $\mu_\pm$ are close to each other. Remarkably, we find that the result is valid generally beyond the latter approximation.

Interestingly, \eqref{Eq:double-scale-zLR-yLR} implies that the conserved charges $e_{L,R,B}$ and $\lambda_{L,R,B}$ in each of the three geodetic segments constituting the full HRT surface assume \textit{constant} values in the double scaling limit \eqref{Eq:double-scale} (although $y_{L,R}^*$ is time-dependent.) Together with the result for linear growth of entanglement entropy, we can conclude that the double scaling limit \eqref{Eq:double-scale} is similar to a steady state regime. 

\subsection{Approach to equilibrium}
For transitions between arbitrary rotating BTZ spacetimes, we find that the entanglement entropy of an interval of length $l$ reaches the final thermal value  at a time $u=\frac{l}{2}$. At $u=\frac{l}{2}$ the geodesic is entirely in the final geometry and grazes the shock ($u=0$) at a single point whose $y$ and $z$ coordinates are given by
\begin{equation}\label{Eq:z-y-converge}
    z_* = \frac{1}{\left(\mu_+^f \coth(\mu_+^f l) +\mu_-^f \coth(\mu_-^f l)\right)}, \quad y_* =\frac{l}{2}.
\end{equation}
These results were obtained earlier in \cite{Kibe:2021qjy} with the assumption that the change in $\mu_\pm$ is small. However, we find that the results hold exactly even beyond the latter approximation. 

We also find numerically that the entanglement entropy in the late time regime $\mu_{\pm}^{i,f}(\frac{l}{2}-u)\ll 1$ behaves as 
\begin{equation}\label{Eq:Exp-Sat}
  S_{\rm EE}^f- S_{\rm EE}(u) \sim \left(\frac{l}{2}-u\right)^{\frac32},
\end{equation}
where $S_{EE}^f$ denotes the final thermal value of the entanglement entropy. See Fig.~\ref{Fig:qnechom-entropy} (b) for an illustration. The $\frac 32$ saturation exponent was seen for transitions between non-rotating BTZ spacetimes in \cite{Liu:2013iza,Liu:2013qca,Hubeny:2013hz}. This result was also obtained analytically in \cite{Kibe:2021qjy} with the assumption that the change in $\mu_\pm$ is small. We again find that the results hold exactly even beyond the latter approximation. 


\section{Generalized Clausius inequalities from QNEC}\label{sec:qnechom-qnecbounds}

In order to obtain the generalized Clausius inequalities, we need to verify the validity of the QNEC by explicitly computing $\mathcal{Q}_\pm$ defined in \eqref{Eq:Qpm-def}. The QNEC implies that $\mathcal{Q}_\pm  \geq 0$ should hold with the null derivatives of the entanglement entropy computed for an entangling interval with arbitrary length $l$. Since the HRT surface does not intersect the shock before the quench (i.e. for $u<0$) and after the thermalization of the entanglement entropy (at $u=l/2$), the QNEC is saturated for both $u<0$ and $u>l/2$ as should be the case (c.f. Sec. \ref{sec:Introduction} and below) in thermal states carrying uniform momentum density which are dual to a BTZ black holes. Therefore, we need to check if $\mathcal{Q}_\pm  \geq 0$ is satisfied in the time period $0\leq u \leq l/2$ for all lengths $l$.

The QNEC requires us to compute the variations of the entanglement entropy of the boundary interval under null deformations of one endpoint with the other endpoint held fixed. Recall that without loss of generality, we take the endpoints of our interval of length $l$ to be $p_L=(0,u,0)$ and $p_R=(0,u,l)$ utilizing spatial translation symmetry. Therefore, we consider the following null deformations of the left end-point 
\begin{equation}\label{Eq:pLdisp}
p_L = \lb 0,u+\frac{\delta x^\pm}{2},\pm\frac{\delta x^\pm}{2} \rb,
\end{equation}
where $\delta x^\pm$ are the amounts of the deformations along the two future oriented $\pm$ null directions, while keeping $p_R$ fixed.  Note that it is sufficient to check the validity of the QNEC at $p_L$ as $\mathcal{Q_\pm}$ at $p_L$ should coincide with $\mathcal{Q_\mp}$ of $p_R$ provided we interchange $\mu_+$ with $\mu_-$ in \textit{both} initial and final states. To see this, we need to utilize the spatial translation symmetry and also $y\rightarrow -y$ reflection symmetry of the BTZ-Vaidya geometries with the latter holding only when $\mu_+$ and $\mu_-$ are also interchanged for \textit{both} initial and final states. Note also that we are considering arbitrary values of $\mu_+$ and $\mu_-$ in the initial and final states.

As discussed in the previous section, we have solved for $y_{L,R}^*$ and $z_{L,R}^*$, the $y$ and $z$ coordinates of the intersection points, respectively using the methodology of Sec. \ref{Sec:qnechom-geodesic-details} for the undeformed boundary end points and have obtained how the entanglement entropy evolves after the quench. To compute $\mathcal{Q}_\pm$, we need to solve for how $y_{L,R}^*$ and $z_{L,R}^*$ change at linear and quadratic orders in $\delta x^\pm$ when the left boundary endpoint $p_L$ undergoes the null deformations given by $\delta x^\pm$, with the right boundary endpoint $p_R$ fixed as mentioned above. Subsequently, expanding the lengths of the three intervals, $L$, $R$ and $B$ in power series of $\delta x^\pm$, we can readily compute the change in entanglement entropy numerically and obtain the derivatives $\partial_\pm S_{\rm EE}$ and $\partial_\pm^2 S_{\rm EE}$ which are needed to compute $\mathcal{Q}_\pm$. 

We recall from the previous section that explicit analytic expressions for $y_{L,R}^*$ and $z_{L,R}^*$ can be obtained at early time perturbatively as a Taylor series of the time $u$ and also in the steady state regime given by the double scaling limit \eqref{Eq:double-scale} for the undeformed left endpoint. It follows that we can also obtain the analytic expressions for $\mathcal{Q}_\pm$ in these two regimes. Numerical solutions for $\mathcal{Q}_\pm$ can of course be obtained generally for arbitrary length $l$ of the entangling interval and at an arbitrary time. We will restrict our discussion here to cases where $\mu_\pm^f \geq \mu_\pm^i$ so that the classical null energy condition is always satisfied at the null shock. It is sufficient to restrict to these cases as the QNEC in the dual CFT is always violated if it is otherwise. 

\begin{figure}
\centering     
\subfigure[]{\includegraphics[scale=0.8]{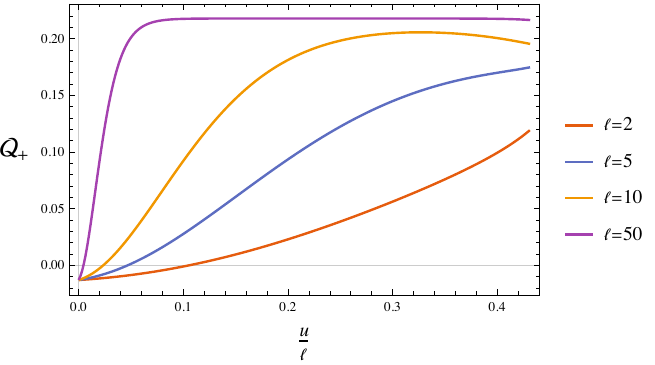}}
\subfigure[]{\includegraphics[scale=0.8]{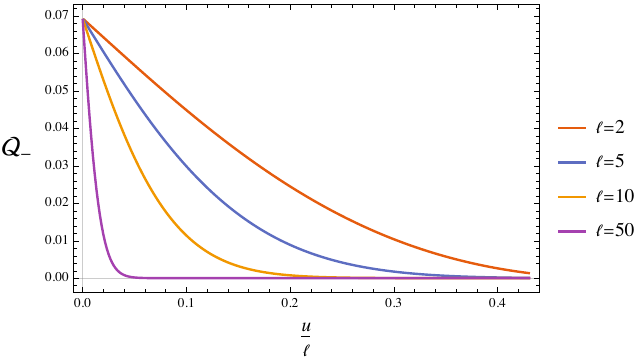}}
\caption{Plots of the evolution of $\frac{6}{c}\cQ_\pm$ for boundary subregions of various lengths indicated on the plots for $\mu_-^i=0.25, \mu_+^i =0.75, \mu_-^f= 0.4, \mu_+^f=0.76$ {(a)} $\cQ_+$. (b) $\cQ_-$.}
\label{Fig:qnechom-qnec-l-dep}
\end{figure}

Representative behavior of $\cQ_\pm$ is shown in Fig.~\ref{Fig:qnechom-qnec-l-dep} for various values of the length $l$ of the entangling interval. As shown in Fig.~\ref{Fig:qnechom-qnec-l-dep} (a) and (b), $\cQ_\pm$ assume non-trivial finite values at $u=0$ although they vanish for $u<0$. We can compute $\cQ_\pm(u=0^+)$ analytically, and as discussed below and evident from Fig.~\ref{Fig:qnechom-qnec-l-dep} (a) and (b), $\cQ_\pm(u=0^+)$ are also independent of the length $l$ of the entangling interval. The analytic values of $\cQ_\pm(u=0^+)$ will be the key in establishing the generalized Clausius inequalities discussed in this section. Note that the choices of $\mu_\pm^i$ and $\mu_\pm^f$ in Fig.~\ref{Fig:qnechom-qnec-l-dep} that are mentioned in the caption correspond to a physically disallowed process as $\cQ_+(u=0)<0$ although $\cQ_-(u=0)>0$.

For $u>0$, $\cQ_+$ behaves non-monotonically in general but never goes below $\cQ_+(u=0)$ (i.e. its value at $u=0$) for $0<u<l/2$ as shown in Fig.~\ref{Fig:qnechom-qnec-l-dep} (a). So the strongest bound from $\cQ_+\geq 0$ is obtained just from the general ($l$ independent) analytic expression for $\cQ_+(u=0)$. At $u=l/2$, however, $\cQ_+$ has a delta function divergence as discussed later. The coefficient of this divergence is always positive when the classical Clausius inequality $\mu_\pm^f \geq \mu_\pm^i$ is satisfied in both left and right moving sectors. Therefore, we do not obtain any non-trivial generalization of the Clausius inequalities from the evaluation of $\cQ_+$ at $u=l/2$ as well.

\begin{figure}
    \centering
    \includegraphics[scale=.8]{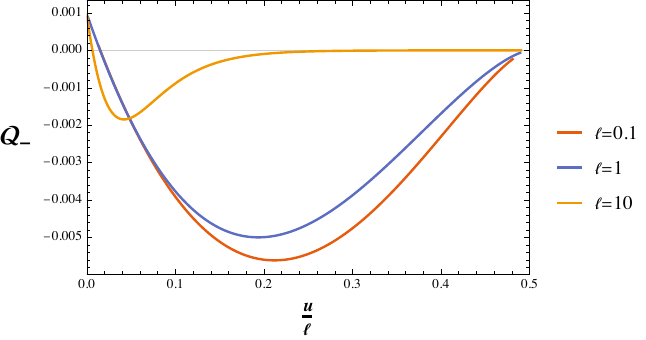}
    \caption{Non-monotonic behaviour of $\frac{6}{c}\cQ_-$ and violation at finite $u$. The curve is plotted with $\mu_-^i=0.25, \mu_+^i =0.75, \mu_-^f= 0.31, \mu_+^f=0.812$.}
    \label{Fig:qnechom-qnecm-violation}
\end{figure}

The behavior of $\cQ_-$ depends on the choices of $\mu_\pm^i$ and $\mu_\pm^f$. We note from Fig.~\ref{Fig:qnechom-qnec-l-dep} (b) that  $\cQ_-$ stays positive for $u\geq 0$ for the choices of $\mu_\pm^i$ and $\mu_\pm^f$ mentioned in the caption of the figure. However, as shown in Fig.~\ref{Fig:qnechom-qnecm-violation}, $\cQ_-$ can be negative for $u>0$ even though it is positive at $u=0$. So to check $\cQ_-\geq 0$, we need to evaluate $\cQ_-$ not only at $u=0$ but also at later times. Unlike the case of $\cQ_+\geq 0$, it is also necessary to examine $\cQ_-\geq 0$ for all lengths $l$ of the entangling interval as the amount of violation of this condition is dependent on $l$ as evident from Fig.~\ref{Fig:qnechom-qnecm-violation}. Nevertheless, as discussed later in this section, we can obtain the strongest bounds from the condition $\cQ_-\geq 0$ by considering the limit $l\rightarrow 0$. We also note both from Fig.~\ref{Fig:qnechom-qnec-l-dep} (b) and Fig.~\ref{Fig:qnechom-qnecm-violation} that $\cQ_- \rightarrow 0$ as $u$ approaches $l/2$ irrespectively of whether the transition is allowed or disallowed.

Note that although $\mu_+^{f(i)}>\mu_-^{f(i)}$ in Fig.~\ref{Fig:qnechom-qnec-l-dep} and ~\ref{Fig:qnechom-qnecm-violation}, the conclusions above are valid for general $\mu_\pm$. The asymmetry in the behavior of $\cQ_\pm$ is simply because when we evaluate $\cQ_\pm$ at $p_L$, the entangling interval contracts under the $+$ deformation and expands under the $-$ deformation. The behavior of $\cQ_\pm$ is therefore interchanged when they are evaluated at $p_R$.

.

\subsection{Examining the QNEC just after the quench}\label{Sec:Qpmu0}
As in the case of the entanglement entropy, we can compute $\cQ_\pm$ at early times $0\leq \mu_\pm^{i,f} u \ll1$ for arbitrary lengths $l$ of the entangling interval as a Taylor series in $u$. At $u=0$, we find that $\cQ_\pm$ are independent of the length $l$ and are explicitly
\begin{align}\label{Eq:qnechom-qnecu0}
  \cQ_+(u=0) = \cQ_+^0 =&\ \frac{c}{6}\frac14\lb 3{\mu_{+}^f}^{2} - {\mu_-^f}^{2} - 3{\mu_+^i}^{2} + {\mu_-^i}^{2} \rb, \nonumber\\
   \cQ_-(u=0) = \cQ_-^0 =&\ \frac{c}{6}\frac14\lb 3{\mu_-^f}^{2} - {\mu_+^f}^{2} - 3{\mu_-^i}^{2} + {\mu_+^i}^{2} \rb.
\end{align}
For $u>0$, the sub-leading linear corrections in $u$ do depend on the length. Explicitly, in the large length limit $ \mu_\pm^{i,f}l\rightarrow\infty$ and for $0\leq \mu_\pm^{i,f} u \ll1$, we obtain
\begin{align}\label{Eq:QNEC-expansion}
  \cQ_+ =&\ \cQ_+^0 + \frac{c}{6}\frac u4\lb {\mu_-^i} \left(3 {\mu_+^f}^2 - 3{\mu_+^i}^2 - {\mu_-^f}^2\right) + {\mu_+^i}\left( 5{\mu_+^f}^2 - 5{\mu_+^i}^2 + {\mu_-^f}^2 \right) + {\mu_-^i}^3 - {\mu_+^i} {\mu_-^i}^2 \rb 
  \nonumber\\ &+\mathcal{O}(u^2),\nonumber \\
  \cQ_- =&\ \cQ_-^0 + \frac{c}{6}\frac u4\lb \mu_-^i \left(-5 {\mu_-^f}^2-{\mu_+^f}^2+{\mu_+^i}^2\right)+\mu_+^i \left(-3 {\mu_-^f}^2+{\mu_+^f}^2-{\mu_+^i}^2\right)+5 {\mu_-^i}^3+3 \mu_+^i {\mu_-^i}^2 \rb  \nonumber\\& +\mathcal{O}(u^2).
\end{align}
We can also obtain the correction to linear order in $u$ for arbitrary lengths. However, we don't report the most general, $l$-dependent linear in $u$ corrections to the QNEC here, since the expressions are rather large.

Note that in case of a transition from vacuum to a non-rotating BTZ with $\mu_\pm^i=0$ and $\mu_\pm^f=\mu$ we get
\begin{equation}
    \cQ_\pm^0 = \frac{c}{6}\frac12 \mu^2 +\mathcal{O}(u^2),
\end{equation}
which reproduces the \textit{half-saturation} of QNEC seen in \cite{Ecker:2019ocp}. 

Furthermore, taking the non rotating limit of the QNEC expressions \eqref{Eq:qnechom-qnecu0} by imposing $\mu_+^{f(i)}=\mu_-^{f(i)} = \mu^{f(i)}$, we simply recover the classical null energy condition  in the dual theory as
\begin{equation}
    \cQ_\pm^0=\frac{c}{6}\frac12\lb {\mu^f}^2-{\mu^i}^2 \rb \geq0.
\end{equation}
The above is equivalent to imposing the classical Clausius inequality $\mu^f \geq \mu^i$ and this has been already implied via the validity of the classical NEC at the bulk shock. Therefore, nothing beyond the usual classical Clausius inequality is obtained in absence of finite momentum densities in both the initial and final states. This is true not only when $\cQ_\pm$ are evaluated at $u=0$ but also at any other time.


Expanding Eq. \eqref{Eq:qnechom-qnecu0} in powers of $\mu_{\pm}^{f}-\mu_{\pm}^i$ we reproduce the QNEC bound obtained in \cite{Kibe:2021qjy} to linear order in $\mu_{\pm}^{f}-\mu_{\pm}^i$.  Furthermore, our analytic expressions in \eqref{Eq:qnechom-qnecu0} and also the expansions as in \eqref{Eq:QNEC-expansion} considered for arbitrary $l$ for $\cQ_\pm$ match with the numerical values obtained in Fig.~\ref{Fig:qnechom-qnec-l-dep}, etc.

\subsection{The double scaling limit}\label{Sec:double-scale}
We recall that the intermediate time behavior for sufficiently large $l$ is very well captured by the double scaling limit \eqref{Eq:double-scale} in which we can solve for the $y$ and $z$ coordinates of the intersection points of the HRT surface and the shock \textit{exactly}. We also recall that in this regime the entanglement entropy grows linearly with the coefficient indicating thermalization proceeding from the two ends of the entangling interval at the speed of light. 

In this limit, the solutions for the $y$ and $z$ coordinates of the intersection points to quadratic order in the null deformation $\delta x^+$ and/or $\delta x^-$ correcting \eqref{Eq:double-scale-zLR-yLR} are
\begin{align}\label{Eq:intpointsxpm}
    z^*_L &=\frac{2}{\mu_-^f+ \mu_+^i+\sqrt{2({\mu_-^i}^2+{\mu_+^f}^2) -(\mu_-^f-\mu_+^i)^2}} +\mathcal{O}({\delta x^\pm}^3),\nonumber\\
    y^*_L &=u +\frac{1}{2 \mu_+^f} \log \left[\frac{\mu_-^i- \mu_+^f+\sqrt{2({\mu_-^i}^2+{\mu_+^f}^2) -(\mu_-^f-\mu_+^i)^2}}{\mu_-^i+ \mu_+^f+\sqrt{2({\mu_-^i}^2+{\mu_+^f}^2) -(\mu_-^f-\mu_+^i)^2}} \right] + \delta x^+ +\mathcal{O}({\delta x^\pm}^3),\nonumber\\
    z^*_2 &=\frac{2}{\mu_+^f+ \mu_-^i+\sqrt{2({\mu_+^i}^2+{\mu_-^f}^2) -(\mu_+^f-\mu_-^i)^2}}+\mathcal{O}({\delta x^\pm}^3),\nonumber\\
    y^*_2&=l -u +\frac{1}{2 \mu_-^f} \log \left[\frac{\mu_+^i+ \mu_-^f+\sqrt{2({\mu_+^i}^2+{\mu_-^f}^2) -(\mu_+^f-\mu_-^i)^2}}{\mu_+^i- \mu_-^f+\sqrt{2({\mu_+^i}^2+{\mu_-^f}^2) -(\mu_+^f-\mu_-^i)^2}} \right]+\mathcal{O}({\delta x^\pm}^3).
\end{align}
assuming $\mu_+^f>\mu_-^f$. From here we can deduce that $\partial_- S_{\rm EE}$, $\partial_-^2 S_{\rm EE}$, $\partial_+\partial_-S_{\rm EE}$ and $\partial_+^2 S_{\rm EE}$ vanish in the double scaling limit whereas $\partial_+ S_{\rm EE}$ is non-vanishing. It also follows that in this limit
\begin{equation}\label{eq:qnechom-qnecplateau}
    \cQ_+ = (s^f -s^i)( \vert\mu_+^f -\mu_-^f\vert+\mu_+^i + \mu_-^i ), \quad \cQ_- = 0.
\end{equation}
Remarkably, the above result also agrees with that obtained in \cite{Kibe:2021qjy} where only small changes in $\mu_\pm$ resulting from the quench was assumed.

As indicated in Fig.~\ref{Fig:qnechom-qnec-l-dep} (a), $\cQ_+$ reaches a plateau at intermediate times for large entangling lengths. We have verified that the numerical answer for the QNEC plateau illustrated in Fig.~\ref{Fig:qnechom-qnec-l-dep} (a) matches this analytic result in \eqref{eq:qnechom-qnecplateau} to a very large accuracy. Furthermore, we note from Fig.~\ref{Fig:qnechom-qnec-l-dep} (b) and Fig.~\ref{Fig:qnechom-qnecm-violation} that for large entangling lengths $\cQ_-$ vanishes at intermediate times to a very good accuracy as predicted from \eqref{eq:qnechom-qnecplateau}. Thus the double scaling limit not only captures analytically the intermediate time behavior of the entanglement entropy, but also the behavior of $\cQ_\pm$ at intermediate times for large entangling lengths to a high accuracy.

\begin{figure}
    \centering
    \subfigure{\includegraphics[scale=0.55]{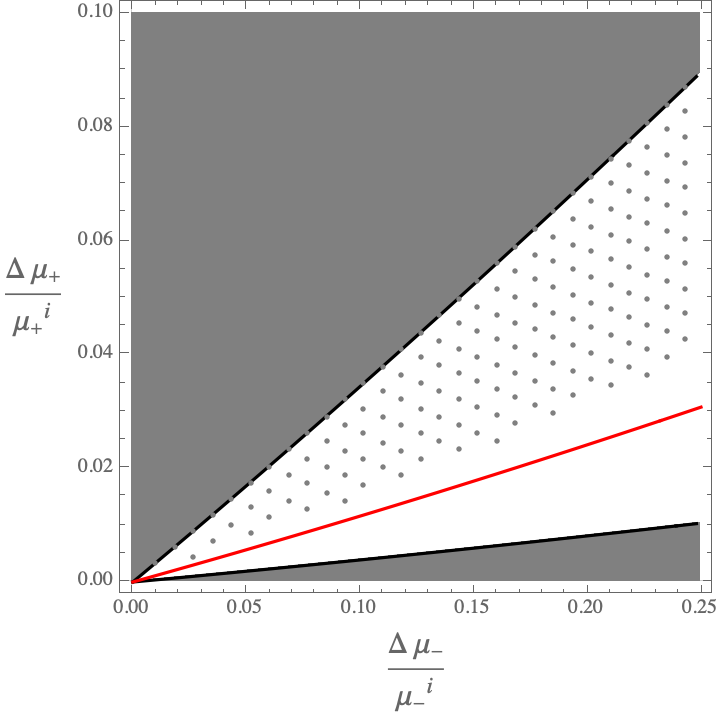}}~~~~
    \subfigure{\includegraphics[scale=0.55]{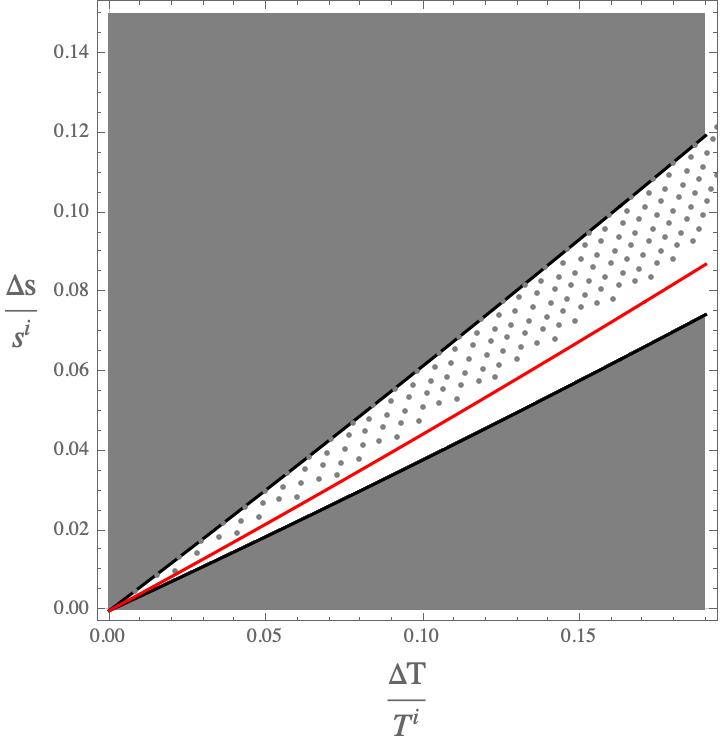}}
    \caption{The parameter space allowed by QNEC is indicated in white. Solid black lines indicate bounds from $u=0$ (obtained analytically) \eqref{Eq:qnechom-qnecu0}. Grey dots indicate the parameters disallowed due to a $\cQ_-$ violation at later times (obtained numerically).\\(a) Bounds on the parameters $\Delta \mu_\pm = \mu_\pm^f-\mu_\pm^i$.  (b) Bounds on the change in the entropy density $\Delta s = s^f-s^i$ and the change in the temperature $\Delta T = T^f-T^i$. The red curve indicates $\Delta J=0$. The initial values are chosen here to be $\mu_+^i=3/4,\mu_-^i=1/4.$}
    \label{Fig:qnechom-gray-region-mu}
\end{figure}

\subsection{Algorithm for determining the allowed transitions}

The evaluation of $\cQ_\pm$ at the moment of quench $u=0$ in Eq. \eqref{Eq:qnechom-qnecu0} itself implies generalization of the Clausius inequality. In Fig.~\ref{Fig:qnechom-gray-region-mu} (a), the possible final states ($\mu_\pm^f$) satisfying the classical Clausius inequalities $\mu_\pm^f\geq \mu_\pm^i$ are shown for a specific choice of the initial state ($\mu_\pm^i$) as mentioned in the caption. The solid black lines in Fig.~\ref{Fig:qnechom-gray-region-mu} (a) indicate where the two inequalities in Eq. \eqref{Eq:qnechom-qnecu0} are saturated. The upper (lower) black line saturates $\cQ_-^0\geq 0$ ($\cQ_+^0\geq 0$). In Fig.~\ref{Fig:qnechom-gray-region-mu} (b), these results have been translated to indicate the allowed changes in entropy and temperature as a result of the quench with the upper (lower) solid black line saturating $\cQ_-^0\geq 0$ ($\cQ_-^0\geq 0$). In both cases, the allowed final states should be in the region bounded by the solid black lines where both inequalities $\cQ_\pm^0 \geq 0$ are satisfied. We readily note that for a fixed change in temperature $\Delta T$, the change in the entropy is bounded both from below and above implying generalization of the Clausius inequality for the quench. 

\begin{figure}
    \centering
    \subfigure[]{\includegraphics[scale=0.65]{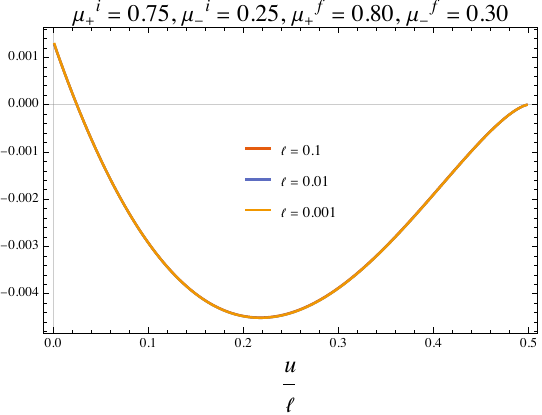}}
    \subfigure[]{\includegraphics[scale=0.65]{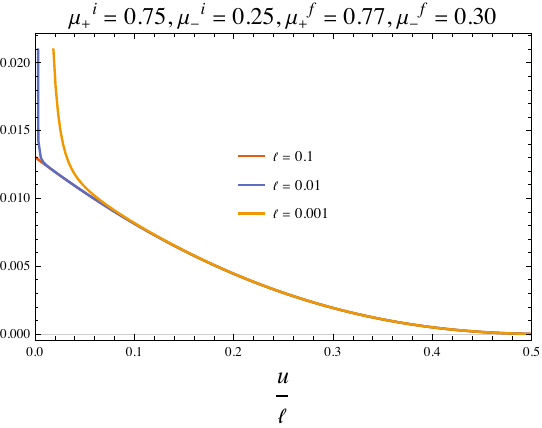}}
    \caption{Plot of $\frac{6}{c} \cQ_-$ for different values of boundary interval lengths for: (a) parameters disallowed by QNEC for $u>0$ and (b) allowed by QNEC. In (a) the plots for $l=0.1,0.01,0.001$ are coincident.}
    \label{fig:small-len-qnec}
\end{figure}

As discussed earlier, although it is sufficient to look at $\cQ_+$ at $u=0$, we must examine $\cQ_-$ for finite times and lengths to determine the strongest constraints on the possible final states imposed by the QNEC. The algorithm to determine the stronger bounds from $\cQ_-$ is as follows.
\begin{enumerate}
    \item  Let us call the upper solid black line where $\cQ_-^0 =0$ as $B^u$ and the lower one where $\cQ_+^0 =0$ as $B^l$. Let the region bounded by $B^u$ and $B^l$ be $R$. We consider lines of fixed $\mu_+^f$ starting from $B^u$ moving inside $R$ along the direction in which $\mu_-^f$ increases.
    \item On $B^u$, we find $\cQ_-\geq 0$ is violated at a finite time ($0<u< l/2$) for any entangling length $l$. As we move inside $R$ to larger values of $\mu_-^f$ for a fixed $\mu_+^f$, we generate data for $\cQ_-(u)$ for $0\leq u\leq l/2$ and for various $l$. For a sufficiently large value of $l$ (e.g. $l=50$ for the choice of initial state in Fig.~\ref{Fig:qnechom-gray-region-mu}), we find that the $\cQ_-(u)$ converges uniformly to the $l\rightarrow\infty$ limit. 
       \item We keep generating data for $\cQ_-(u)$ for smaller $l$ (keeping $\mu_\pm^f$ fixed). We find that inside $R$ and for sufficiently small length $l$, 
       \begin{equation}\label{Eq:Scale-Qm}
        Q_-(u; l,\mu_\pm^i,\mu_\pm^f) \geq F(u/l;\mu_\pm^i,\mu_\pm^f)        
        \end{equation}   
holds for $0\leq u/l < 1/2$, i.e. as we decrease $l$, $\cQ_-(u)$ is bounded by a function of $u/l$ without any explicit $l$ dependence for fixed initial and final states. This feature has been illustrated in Fig. \ref{fig:small-len-qnec} where we notice that it holds irrespectively of whether $\cQ\geq 0$ is violated or not. In fact, we note from Fig. \ref{fig:small-len-qnec} (a) that when $\cQ_-(u)\geq 0$ \textit{is} violated, $\cQ_-(u)$ actually \textit{converges} to $F(u/l)$ instead of being bounded from below by $F(u/l)$, and furthermore             \begin{equation}
       F(x;\mu_\pm^i,\mu_\pm^f) = \lim_{l\rightarrow 0}Q_-(u=xl; l,\mu_\pm^i,\mu_\pm^f).
       \end{equation}
 On the other hand, we note from Fig. \ref{fig:small-len-qnec} (b) that when $\cQ_-(u)\geq 0$ is \textit{not} violated, then
 \begin{equation}
       F(x;\mu_\pm^i,\mu_\pm^f) = \lim_{l\rightarrow l_*(\mu_\pm^i,\mu_\pm^f)}Q_-(u=xl; l,\mu_\pm^i,\mu_\pm^f).
       \end{equation}
and therefore it is sufficient to check whether $\cQ_-(u)\geq 0$ is satisfied for $l\geq  l_*(\mu_\pm^i,\mu_\pm^f)$.
       \item For each final state, we keep decreasing $l$ until we find violation of $\cQ_-(u)\geq 0$ or until $l$ takes the value $ l_*(\mu_\pm^i,\mu_\pm^f)$. In the latter case, we can be sure that the QNEC is satisfied in the quench for that final state, and therefore it is an allowed physical process.
       \end{enumerate}
 
 In Fig.~\ref{Fig:qnechom-gray-region-mu} (a), the allowed final states obtained via the above algorithm form the white region. Note that the white region is a subregion of $R$. The intersection of the complement of the white region and $R$ is the dotted region in Fig.~\ref{Fig:qnechom-gray-region-mu} where $\cQ_-(u)\geq0$ is violated at a finite time after the quench. For the choice of initial state in this figure, $l_*$ is typically $0.1$, and therefore it is sufficient to check the QNEC for $l$ lying in the range $0.1\leq l\leq 50$ as discussed above. 
 
 These bounds are qualitatively similar to the results obtained in \cite{Kibe:2021qjy} assuming that the final states were not very different from the initial state. The results of \cite{Kibe:2021qjy} apply only very close to the origin of Fig.~\ref{Fig:qnechom-gray-region-mu}.
 
 When our results are converted to region plots of the change in entropy and temperature as a result of the quench as in Fig.~\ref{Fig:qnechom-gray-region-mu} (b), we obtain the generalized Clausius inequalities like those which have been established for finite dimensional systems using distance measures in Hilbert space \cite{PhysRevLett.105.170402,PhysRevLett.107.140404,Van_2021,RevModPhys.93.035008}. 

 Finally, we find that transitions in which the total momentum density does not change, i.e. for which 
 \begin{equation}
     \Delta j = 0 \implies {\mu_+^f}^2 - {\mu_-^f}^2 = {\mu_+^i}^2 - {\mu_-^i}^2
 \end{equation}
are always allowed. The curve $\Delta j =0$ has been shown in red in Fig.~\ref{Fig:qnechom-gray-region-mu}, and we note that it is in the white region.


\subsection{A short note for $u=l/2$}

At $u=l/2$, $\cQ_+$ diverges as $\delta(u-l/2)$ while $\cQ_-$ vanishes. Note that the times very close to $u=\frac{l}{2}$ have been excised from all plots of $\cQ_\pm$ to avoid the (possible) delta function divergences. Such a divergence in QNEC is expected whenever the entanglement entropy is continuous but has a kink \cite{Ecker:2020gnw}. It is easy to see that $\partial_u S_{\rm EE}$ is discontinuous at $u=\frac l2$ as implied by \eqref{Eq:Exp-Sat}. The coefficient of the delta function can be obtained as follows. 

Let us consider the entanglement entropy $S_{\rm EE}(x^+)$ with $p_L$ displaced to
\begin{equation}\label{Eq:pLdisp2}
p_L = \lb 0,\frac{l}{2}+\frac{ x^+}{2},\frac{x^+}{2} \rb,
\end{equation}
 in the $(z,u,y)$ coordinates with $p_R$ fixed. Unlike \eqref{Eq:pLdisp}, the null displacement of $p_L$ is finite and not infinitesimal. Crucially for this displacement, the $x^-$ coordinate of $p_L$ remains fixed at $l/2$ while the $x^+$ coordinate of $p_L$ is $l/2 + x^+$. Then we should have
\begin{equation}
    S_{\rm EE} = \Theta \lb -x^+ \rb S^{i}(x^+) +\Theta\lb x^+\rb S^f(x^+),
\end{equation}
where $S^f$ is the final thermal state result and $S^i$ is the entropy before thermalization. Note $S_{\rm EE}$ is continuous at $u=l/2$ but its derivative is not. Therefore, differentiating the above expression with respect to the null coordinate $x^+$ once leads to a vanishing result, but differentiating twice we get a delta function divergence. We can readily see that
\begin{equation}
    \mathcal{Q}_+\left(u=\frac l2\right) = -\delta(x^+)\Delta\partial_+S_{\rm EE}+ \text{finite terms},
\end{equation}
where $\Delta \partial_+S_{EE}= \partial_+S^f -\partial_+ S^i$ and the finite terms have been already analyzed in this section. Numerically, we find that $\Delta\partial_+S_{\rm EE}$ is negative whenever $\mu_\pm^f \geq \mu_\pm^i$ and therefore we do not obtain anything beyond the classical Clausius inequalities by requiring that $\cQ_+\geq 0$ at $u=l/2$. A similar study with the other null deformation gives us that $\cQ_-$ vanishes at $u=l/2$.

    \label{Eq:deltafncoef}

\section{Recovering information of the initial state}\label{sec:recover}
An interesting question to explore is the amount of information of the initial state retained by the post-quench density matrix. This can be probed using the post-quench entanglement entropy and Renyi entropies. Furthermore, this question is likely also relevant for experiments where techniques for an indirect measurement of the entanglement spectrum have been developed \cite{Klich:2008un, Islam:2015mom}. For this purpose, one can use spacelike intervals of various lengths such that all points in the interval are at equal time in different frames including those which are boosted with respect the the frame of the intial/final thermal state. Equivalently, one can also study how the entanglement entropies change under small displacements of one of the endpoints of a spatial interval in the frame comoving with the initial/final thermal state, i.e. study $\partial_\pm S_{\rm EE}$, and $\partial_+^2 S_{\rm EE}$, $\partial_+\partial_- S_{\rm EE}$, etc. at various times after the quench.

\subsection{Recovering initial state just after the quench}

As discussed in Sec. \ref{Sec:Qpmu0}, we can compute all derivatives of $S_{\rm EE}$ measuring change of the entanglement under displacement of one of the endpoints analytically as a Taylor series in time just after the quench at $u=0$. In Sec. \ref{Sec:Qpmu0}, we reported the results for $\cQ_\pm(u)$. The first derivatives of  $S_{\rm EE}$ turn out to be
\begin{equation}\label{Eq:FDu0}
    \partial_{\pm} S = \mp\frac{c}{6} \mu_\pm^i \coth(l \mu_\pm^i),
\end{equation}
while the second derivatives of $S_{\rm EE}$ turn out to be
\begin{align}\label{Eq:SDu0}
\partial_{+}\partial_{-} S_{\rm EE} &= \frac{c}{24}\left({\mu_-^f}^2-{\mu_-^i}^2+{\mu_+^f}^2-{\mu_+^i}^2\right).\nonumber\\
    \partial^2_+S_{\rm EE} &=\frac{c}{24} \left({\mu_-^f}^2 - {\mu_-^i}^2+{\mu_+^f}^2 - {\mu_+^i}^2 -4 {\mu_+^i}^2 \csch^2(\mu_+^i l)\right),\nonumber\\
    \partial^2_-S_{\rm EE} &=\frac{c}{24} \left({\mu_-^f}^2 - {\mu_-^i}^2+{\mu_+^f}^2 - {\mu_+^i}^2 -4 {\mu_-^i}^2 \csch^2(\mu_-^i l)\right)
\end{align}
at the time of quench, $u=0$. (Although the $l$ dependence vanishes for $\cQ_\pm(u)$ evaluated at $u=0$ as reported in in Sec. \ref{Sec:Qpmu0}, note that they are present for the values of $\partial_\pm{S}_{EE}$, $\partial_+^2S_{\rm EE}$ and $\partial_-^2S_{\rm EE}$ at $u=0$ individually.) We find that when the left endpoint is displaced, the segment $R$ of the co-dimension two bulk HRT surface connecting the shock to the right endpoint does not contribute to the first and second derivatives of $S_{\rm EE}$ at the time of quench.

It is easy to see that we can recover $\mu_\pm^i$ from $\partial_\pm S_{\rm EE}$ or from $( \partial^2_+ -\partial_{+}\partial_{-})S_{\rm EE}$ and  $( \partial^2_- -\partial_{+}\partial_{-})S_{\rm EE}$ at $u=0$.


\subsection{The double-scaling limit}

In the double-scaling limit \eqref{Eq:double-scale}, the entanglement entropy grows linearly with the rate \eqref{Eq:qnechomo-vs} corresponding to thermalization occurring from both endpoints at the speed of light, $\cQ_+$ plateaus and $\cQ_-$ vanishes as implied by \eqref{eq:qnechom-qnecplateau}. Using the explicit results for the $y$ and $z$ coordinates of the intersection points between the HRT surface and the shock, we can readily obtain $S_{\rm EE}$ and its various derivatives which turn out to be 
\begin{align}
    &S_{\rm EE} = \frac{c}{3} \left(\mu_+^f+\mu_-^f-\mu_+^i-\mu_-^i\right) u,\nonumber\\
    &\partial_{+} S_{\rm EE} =\frac{c}{6}\left(\mu_-^f-\mu_+^i-\mu_-^i\right),\nonumber\\
    &\partial_{-} S_{\rm EE} = 0, \quad \partial^2_{\pm} S_{\rm EE} = \partial_{+}\partial_{-}S_{\rm EE}=0.
\end{align}
The entanglement entropy, and its first and second derivatives depend only on the initial entropy density, $s^i = \frac{c}{6}(\mu_+^i+\mu_-^i)$. Thus it is hard to recover the information of the initial state in this limit. We have not been able to prove that we cannot recover both $\mu_+^i$ and $\mu_-^i$ using higher order derivatives of $S_{\rm EE}$ in the double scaling limit but this indeed could be the case.

Since the double scaling limit captures very accurately the behavior of the entanglement entropy and its derivatives with respect to displacements of one of its endpoints for large intervals $\mu_\pm^{i,f}l \geq 1$ at intermediate times as reported before, we learn that the information of the initial state cannot be easily recovered from the latter data. As the entanglement entropy of smaller intervals thermalize quickly, we also learn that the information recovery of the initial state is difficult generally at late times. To understand more details of the scrambling of information which makes recovery harder, it should be useful to study the Renyi entropies in the double scaling limit and their derivatives under displacements of one of the endpoints ({since these have more fine-grained information than the entanglement entropy}). 

\section{A short note on the quantum dominant energy conditions}\label{Sec:QDEC}
While the quantum null energy condition has been established firmly for a general two-dimensional CFT, it could be useful to examine other energy conditions which have been proven so far in limited contexts. Such a well-known proposal for a two-dimensional QFT \cite{Wall:2017blw} is constituted by the Quantum Dominant Energy Conditions (QDEC), which are
\begin{equation}
    \braket{t_{ab}} v^a w^b \geq \frac{1}{2\pi} v^a w^b \left(\epsilon^c_a \epsilon^d_b \partial_c\partial_d S_{\rm EE}\right),
    \label{Eq:QDEC}
\end{equation}
where the derivatives are with respect to the displacements of one of the end points of the entangling interval, $v$ and $w$ are vectors in the future light cone of that endpoint, and $\epsilon$ is the Levi-Civita tensor. 

We can readily evaluate the QDEC at $u=0$ using \eqref{Eq:SDu0}. We find that the QDEC gives the strongest bounds when both $v$ and $w$ coincide with the $+$ or $-$ null vectors and $l \to \infty$. In this case, the strongest QDEC correspond to weaker version of QNEC obtained by removing the $(\partial_\pm S_{\rm EE})^2$ terms in \eqref{Eq:Qpm-def}. These imply that
\begin{align}
    3 {\mu_+^f}^2-3 {\mu_+^i}^2-{\mu_-^f}^2+{\mu_-^i}^2&\geq0\nonumber\\
    3 {\mu_-^f}^2-3 {\mu_-^i}^2-{\mu_+^f}^2+{\mu_+^i}^2&\geq0.
\end{align}
Note that the above are weaker than \eqref{Eq:qnechom-qnecu0} (i.e. the final states allowed by \eqref{Eq:qnechom-qnecu0} which are obtained from evaluating the QNEC just after the quench are only a subset of those allowed by the above inequalities). The weakest version of QDEC corresponds to $v^a\partial_a$ and $w^a\partial_a$ coinciding with the $\partial_t$ in which case it just amounts to the classical Clausius inequality. Similar conclusions also hold at later times in the double scaling limit. 


\section{Discussion}\label{sec:discuss}
In this work, we have developed an algebraic method for computing the HRT surfaces for quenches leading to transitions between arbitrary quantum equilibrium states in holographic CFTs, and have determined the generalized Clausius inequalities which should hold in transitions between generic thermal states with uniform momentum densities via the validity of the QNEC. Our construction of the Ba\~nados-Vaidya geometries describing such quenches are independent of the details of the dual two-derivative 2+1-D gravitational theory as the bulk energy-momentum tensor localized on the null shock is determined completely by the initial and final quantum equilibrium states. We find that even when the bulk energy-momentum tensor on the null shock satisfies the (classical) NEC, the generalized Clausius inequalities, i.e. the QNEC can be violated in the corresponding quench in the dual field theory. We also find that the transition is always allowed when the change in momentum density vanishes, and the energy density does not decrease in both the left and right moving sectors. The lower and upper bounds on the irreversible entropy production obtained in this work can be converted to lower and upper bounds on the rate of growth of entanglement entropy, e.g. the rate of diffusive growth of entanglement entropy at early time \eqref{Eq:D} derived for arbitrary quenches.

The proof of the QNEC in holographic field theories in \cite{Koeller_2016} shows that if the NEC is satisfied in the bulk of a \textit{smooth} classical geometry which is the solution of a two-derivative gravitational theory, then it should imply that the QNEC is satisfied in the dual field theory. However this does not imply anything for a solution with a null shock unless it is the limit of a smooth solution of a two-derivative gravity theory. When momentum density vanishes at all times, the BTZ-Vaidya geometry with the null shock is in fact a limit of a smooth solution of a massless scalar field coupled to gravity as discussed in Sec.~\ref{Sec:qnechom-quenches}. Therefore, our finding that the QNEC is satisfied when the change in momentum density vanishes, is consistent with the proof. Our general results suggest that the  Ba\~nados-Vaidya geometries where the NEC is satisfied in the bulk but the QNEC is violated in the dual field theory cannot be realized as limits of smooth solutions of a two-derivative gravity theory, i.e. Einstein's gravity minimally coupled to matter fields. 
{Therefore, even if such transitions are possible in \textit{finite} time at large $N$ and strong coupling (so that the dual gravitational theory is Einstein's gravity minimally coupled to matter fields), it should \textit{not} be possible to take the limit in which the transition becomes instantaneous \textit{in the bulk}.}\footnote{{Consider sourcing the dual $1+1$-D CFT with a gravitational potential via $g_{tt} = -1 - e^{-2\phi(x)} f(t)$ with $f(t)$ being a Gaussian function. This leads to a gravitational force acting for finite time leading to the injection of net momentum to the thermal matter. However, even in the limit in which $f(t)$ becomes a delta function, the transition of the system to the final equilibrium with the added uniform momentum density can occur over a period of time. It should be possible to study this explicitly via generalized hydrodynamics \cite{Doyon:2023zcl} which can explain how the right moving excitations attain different momentum from the left movers. It would also be interesting to study this in a holographic setup. Furthermore, we should study examples in which the modification of the boundary metric is not required and understand generally if the transition can be made instantaneously in an appropriate limit when it is not forbidden at large $N$ and strong coupling. We thank the referee for a relevant discussion.}} However, the disallowed transition {with an instantaneous limit in the bulk} may be possible in the full classical string field theory which is the dual gravitational description of the holographic CFT at \textit{finite} coupling and large $N$. In the latter case, the HRT prescription itself should be modified.

Pertinently, it has been shown in \cite{Banerjee:2024sqq} that one can obtain the Nambu-Goto equation from 2+1-D gravity via the junction conditions.\footnote{This result may not generalize to higher dimensions. As shown in \cite{Banerjee:2024sqq}, the independent equations for the junction conditions outnumber the degrees of freedom of the junction in higher dimensions ($D>3$). So, {although some special symmetric solutions of the world-volume extremization equations can be expected to be realized in the limit where the brane tension vanishes, we should not be able to obtain a generic solution of these equations in the tensionless limit for $D > 3$.}} More precisely, the general solutions of 2+1-D gravity with a junction have one-to-one correspondence with the solutions of the non-linear Nambu-Goto equations up to a finite number of rigid parameters related to worldsheet and spacetime isometries. These solutions of 2+1-D gravity are dual to novel types of interfaces in the universal sector of 1+1-D CFTs featuring time-reparametrization at the interface. (In fact, one can decode the bulk Nambu-Goto solution from the time-reparametrization at the interface \cite{Banerjee:2024sqq}). These solutions give a method for constructing a large variety of quantum processors (giving a map between the Hilbert spaces of the CFT on the left and right sides of the interface) utilizing the fundamental string junction in 2+1-D gravity. 

Such solutions of 2+1-D gravity with the fundamental string junction, however, are not {expected to be} limits of smooth solutions of theories where bulk matter fields (with the matter action involving only two derivatives) are coupled minimally to Einstein's gravity. Therefore, we can expect that our methods when extended to such solutions will produce generalized Clausius inequalities via the validity of QNEC constraining quantum processors which can be physically realized via the interfaces dual to fundamental string junctions in 2+1-D gravity. Such results should give us new fundamental insights into quantum gravity. This can be expected as the QNEC itself was conjectured as a way to satisfy the generalized second law in semi-classical gravity \cite{Bousso_2016}.

Another important result of our work is that the initial state can be recovered from the entanglement entropy of different spatial intervals (which are on a constant time slice in the frame of observers who have finite boosts with respect to the final equilibrium state) at early times after the quench but not at late times particularly in the regime where the entanglement entropy thermalizes at the speed of light. {This leads us to pose an interesting question about the scrambling of quantum information in holographic CFTs. Specifically, we can ask if the (time dependent) entanglement spectrum of various (space-like) intervals of a fixed finite length $l$ can be used to recover the initial state at arbitrary times after the quench which is fundamentally a unitary process. It would be interesting to study the time evolution of the Renyi entropies for this purpose as these have more fine grained information about the entanglement spectrum.  This should give us a way to precisely quantify information scrambling in a subsystem. Furthermore, we should also be able to extract how the time-scale of scrambling of information in an interval (subsystem) depends on its length (size).}

On a related theme, it is possible that Renyi entropies are associated with new types of QNECs which should hold in states of arbitrary CFTs \cite{Lashkari:2018nsl}. These Renyi QNECs have been proven for free and super-renormalizable bosons in \cite{Moosa:2020jwt} and fermions in \cite{Roy:2022yzm}. The study of evolution of Renyi entropies in quenches can therefore provide stricter versions of generalized Clausius inequalities in relativistic quantum field theories. Certainly, a pertinent question is which generalized version(s) of the QNEC give the strictest restrictions on physical processes if such generalizations indeed exist for generic and/or holographic CFTs.

{A variety of explicit quench protocols have been studied in CFTs (see \cite{Bernard:2016nci} and references therein).  
A particularly interesting quench protocol involves instantaneously joining two half-line CFTs prepared at different temperatures leading to the formation of  a steady state \cite{Bernard:2016nci}. This setup has also been studied holographically in \cite{Erdmenger:2017gdk}. It would be interesting to study if QNEC imposes any bounds on such quenches leading to the formation of steady states. }

{Finally, from the holographic point of view, it should be interesting to understand the necessary and sufficient conditions (beyond the classical NEC) which the stress tensor localized on a bulk shock or junction should satisfy generally so that the QNEC is also satisfied in the corresponding process in the dual field theory. Such a set of conditions for the bulk stress tensor on the co-dimension one hypersurface can be non-local.}

\section*{Acknowledgments}

We would like to thank Costas Bachas, Giuseppe Policastro, Mukund Rangamani and Ronak Soni for valuable discussions. The research of PR is supported by the NRF Postdoctoral Fellowship. The research of AM has been supported by Fondecyt grant 1240955.

\bibliography{references}
\bibliographystyle{JHEP}

\end{document}